\documentclass{article} 
\usepackage{iclr2024_conference,times}

\usepackage{microtype}
\usepackage{graphicx}
\usepackage{subfig}
\usepackage{bbding}
\usepackage{booktabs} 
\usepackage{multirow}
\usepackage{enumitem}
\setlist[enumerate]{nosep}
\usepackage{bbm}
\usepackage{overpic}

\usepackage{amsmath}
\usepackage{amssymb}
\usepackage{mathtools}
\usepackage{amsthm}
\usepackage{thmtools}

\usepackage{amsmath,amsfonts,bm}

\def\eqref#1{equation~\ref{#1}}

\def\1{\bm{1}}

\DeclareMathAlphabet{\mathsfit}{\encodingdefault}{\sfdefault}{m}{sl}
\SetMathAlphabet{\mathsfit}{bold}{\encodingdefault}{\sfdefault}{bx}{n}

\usepackage{hyperref}
\usepackage{url}

\theoremstyle{plain}
\newtheorem{theorem}{Theorem}[section]

\theoremstyle{definition}
\newtheorem{definition}[theorem]{Definition}

\theoremstyle{remark}

\newcommand{\ours}{\textsc{PROflow}}

\newcommand{\RR}{\mathbb{R}}

\newcommand{\xP}{{\mathbf{x}_p}}
\newcommand{\xE}{{\mathbf{x}_e}}
\newcommand{\xlinker}{\mathbf{x}_\ell}
\newcommand{\xWe}{{\mathbf{x}_{W_e}}}
\newcommand{\xWp}{{\mathbf{x}_{W_p}}}
\newcommand{\nP}{{N_p}}
\newcommand{\nE}{{N_e}}

\newcommand*\diff{\mathop{}\!\mathrm{d}}
\usepackage{colonequals}

\usepackage{wrapfig}

\title{PROflow: An iterative refinement model\\for PROTAC-induced structure prediction}

\author{Bo Qiang \\
Department of Pharmaceutical Science \\
Peking University \\
Beijing, China \\
\texttt{colinqiang@pku.edu.cn} \\
\And
Wenxian Shi \\
Department of Computer Science \\
Massachusetts Institute of Technology \\
Cambridge, MA, USA \\
\texttt{wxsh@mit.edu} \\
\And
Yuxuan Song \\
Institute for Artificial Industry Research \\
Tsinghua University \\
Beijing, China \\
\texttt{yxsong0816@gmail.com} \\
\And
Menghua Wu \\
Department of Computer Science \\
Massachusetts Institute of Technology \\
Cambridge, MA, USA \\
\texttt{rmwu@mit.edu}
}

\iclrfinalcopy 
\begin{document}

\maketitle

\begin{abstract}

Proteolysis targeting chimeras (PROTACs) are
small molecules that trigger the breakdown of traditionally ``undruggable'' proteins by binding simultaneously to their targets and degradation-associated proteins.
A key challenge in their rational design is understanding their structural basis of activity.
Due to the lack of crystal structures (18 in the PDB),
existing PROTAC docking methods have been 
forced to simplify the problem into a distance-constrained protein-protein docking task.
To address the data issue,
we develop a novel pseudo-data generation scheme that requires only binary protein-protein complexes.
This new dataset enables \ours{},
an iterative refinement model
for PROTAC-induced structure prediction that models the full PROTAC flexibility during constrained protein-protein docking.
\ours{} outperforms the state-of-the-art across docking metrics and runtime.
Its inference speed enables the large-scale screening of PROTAC designs, and computed properties of predicted structures achieve statistically significant correlations with published degradation activities.

\end{abstract}
\section{Introduction}

Targeted protein degradation is an emerging paradigm
in rational drug design
that induces the breakdown of ``undruggable'' proteins~\citep{zhao2022targeted}.
Proteolysis targeting chimeras (PROTACs)
are small molecules that
achieve this by simultaneously binding a protein of interest (POI) and a degradation-associated protein (e.g. E3 ligase)~\citep{zou2019protac, hu2022recent}.
In contrast to small molecule drugs, which attach to predefined sites on their protein targets,
PROTACs operate by inducing a stable, \emph{ternary} complex between themselves and two proteins which don't typically interact.
This design task is highly structural, but
modeling these structures has eluded both experimentalists and computationalists. There are only 18 such structures in the PDB~\citep{weng2023protac}, and PROTAC docking algorithms have yet to see widespread use~\citep{troup2020current}, 
as existing search-based methods either are too slow or oversimplify the task.
Finally, while deep learning has shown substantial promise in molecular docking~\citep{corso2022diffdock},
there are currently no end-to-end deep learning methods for PROTAC docking, due to the lack of data.

\begin{wrapfigure}{R}{0.45\textwidth}
\vskip -0.2in
\centering
\includegraphics[width=0.44\columnwidth]{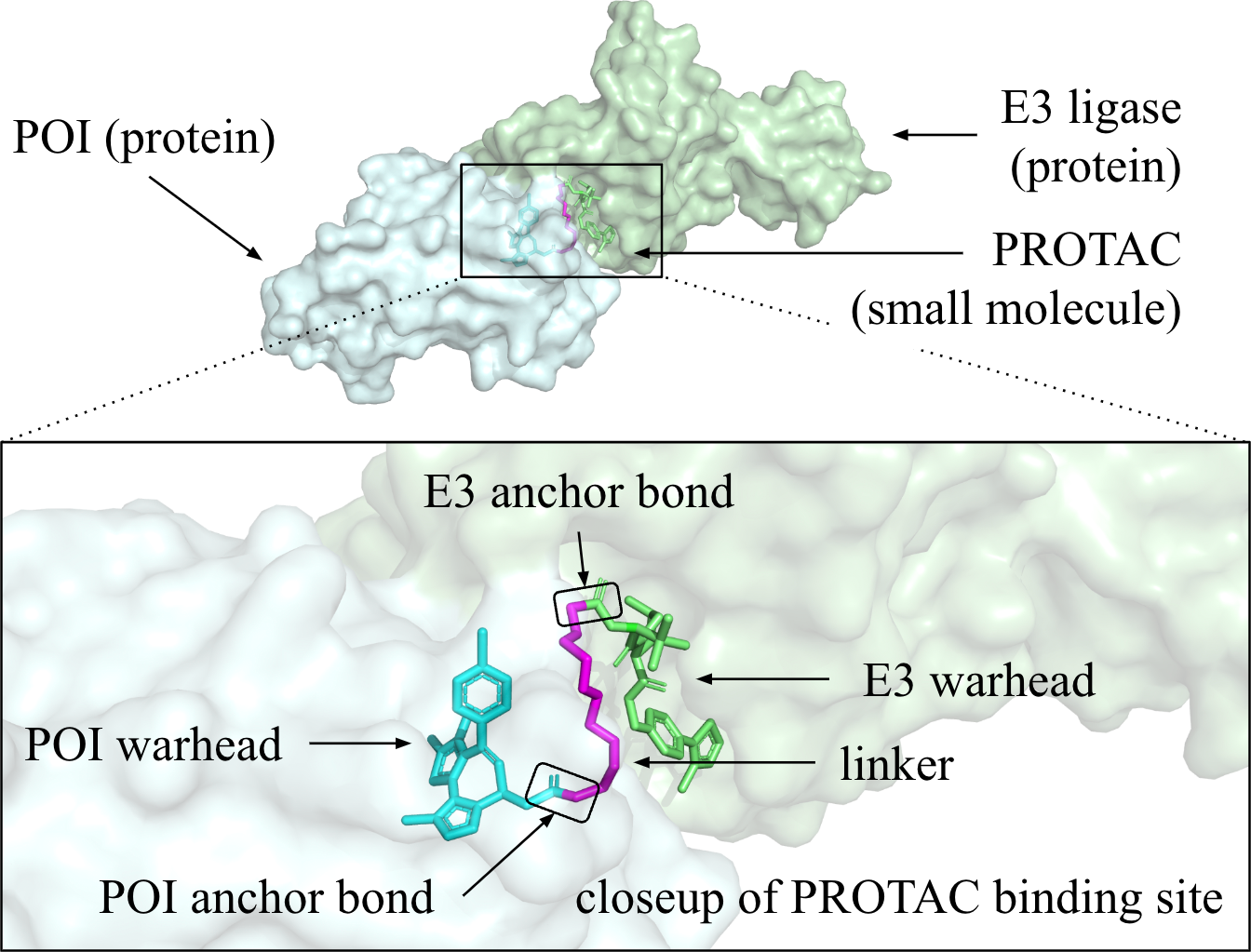}
\caption{PROTACs are small molecules composed of two ``warheads'' and a connecting linker. The warheads bind to the E3 ligase and protein of interest (POI), while the flexible linker brings the two proteins into proximity.
Top: ternary complex between POI (left), E3 ligase (right), PROTAC (small molecule).
Bottom: PROTAC binding site, with respective warheads and anchor bonds.  PDB 5T35.}
\vskip -0.1in
\label{motivation}
\end{wrapfigure}

Specifically, given unbound 3D structures of the POI and E3 ligase and the 2D PROTAC molecular graph, our goal is to predict the bound poses of these three objects (``ternary complex,'' Figure~\ref{motivation}).
To this end, we propose \ours{}, an iterative refinement model for PROTAC-induced structure prediction.
We frame the task as a conditional generation problem, where we learn the distribution over rigid-body protein transformations that respect the existence of a connecting PROTAC linker.
To train this model in the absence of real ternary complexes, we create a pseudo-ternary dataset using a novel data generation scheme that pairs binary protein-protein data with appropriate PROTAC linkers~\citep{townshend2019end, weng2023protac}.

Empirically, \ours{} outperforms existing PROTAC docking methods in predicting PROTAC-induced complexes (8.35 interface RMSD) and E3-POI interfaces (0.264 Fnat).
In addition, \ours{} runs up to 60 times faster than the only alternative that considers full PROTAC structures, enabling the virtual screening of hundreds of designs within 5 hours.
As a direct result, we are able to make predictions over the entire PROTAC-DB~\citep{weng2021protac} and show statistically-significant correlations between Rosetta-computed properties of our structures~\citep{leaver2011rosetta3} and published degradation activity.
In summary, our main contributions are as follows.
\begin{enumerate}
    \item To the best of our knowledge, \ours{} is the first end-to-end deep learning approach
    for PROTAC-induced structure prediction. In contrast to previous work, we consider the
    full conformational landscape of the PROTAC linker during the entire sampling process.
    \item We create and provide a new pseudo-ternary dataset from binary protein-protein complexes and PROTAC linker graphs, to facilitate further development of models for this task.
    \item We achieve state-of-the-art performance on
    PROTAC docking benchmarks with up to 60 times speedup. In addition, computed properties of our predicted complexes
    show statistically significant correlations with published degradation activity,
    highlighting \ours{}'s utility for future design tasks.
\end{enumerate}

\section{Background and related work}
\paragraph{Molecular docking}
Molecular docking plays a pivotal role in structure-based drug design~\citep{morris2008molecular, hwang2010protein}.
Given two or more molecules (e.g. small molecules and/or proteins), the goal is to predict their bound pose.
Traditional docking algorithms break down the problem into two steps: a sampling method enumerates candidate 3D poses, and a scoring function ranks these poses~\citep{trott2010autodock, yan2020hdock}.
More recently, deep learning models have also been developed for molecular docking. These methods formulate
the problem as either a regression~\citep{stark2022equibind, lu2022tankbind, zhang2022e3bind} or a generative task~\citep{corso2022diffdock, ketata2023diffdock}.

\paragraph{PROTAC docking}

Current methods for PROTAC docking
incorporate PROTACs as constraints or filters for protein-protein
docking.
A common approach first samples PROTAC conformations and E3-POI complexes separately;
then (optionally) re-inserts and relaxes the PROTAC linkers to minimize collisions~\citep{drummond2019silico, drummond2020improved}.
This workflow has been used to query
the structural bases of known PROTAC interactions~\citep{weng2021integrative, bai2021rationalizing}.
The PROTAC can also be simplified into a set of distances for constrained protein-protein docking~\citep{schneidman2005patchdock,li2022importance}.
The primary drawback to these approaches
is that they largely ignore the linker's flexibility by reducing
it to a set of distances or fixed conformations during protein-protein sampling.

\paragraph{Flow matching}

Flow matching~\citep{lipman2023flow,albergo2023stochastic} is an iterative refinement, generative modeling framework that has achieved the state-of-the-art across standard generation tasks~\citep{tong2023improving} and biomolecular modeling~\citep{stärk2023harmonic, yim2023fast, song2023equivariant}.
Flow matching has also been extended to Riemannian manifolds~\citep{lipman2024flow}.
In this work, we develop an iterative refinement model based on flow matching, with chemically-informed modifications to ensure structural validity.

The goal of flow matching is to learn a vector field $u_t$ that transports a prior distribution $x_0 \sim q$ to the target data distribution $x_1 \sim p_\text{data}$.
To efficiently learn $u_t$,
we first define a \emph{conditional flow map}
$\psi_t(x_0 \mid x_1)$,
whose derivative with respect to time yields the
\emph{conditional vector field}
$
\label{eq:cfm}
    \frac{d}{dt} \psi_t(x_0 \mid x_1)
    = u_t(x \mid x_1).
$
The conditional vector field gives rise to a
\emph{conditional probability path} $p_t(x\mid x_1),~t\in[0,1]$
via the continuity equation,
where $p_0(x\mid x_1) = q(x)$ (prior) and $p_1(x\mid x_1) \approx \delta(x - x_1)$ (approximate Dirac).

The conditional flow matching loss is
$
    \label{eq:loss_f}
    \mathcal{L}_{\text{CFM}}(\theta)=\mathbb{E}_{t, q(x), p_t(x\mid x_1)} \left\|v_{\theta}(x, t)-u_{t}\left(x \mid x_1 \right)\right\|^2
$
where $v_\theta$ is the output of a neural network $\theta$.
It has been shown that in expectation,
regressing against conditional vector field $u_t(x\mid x_1)$
is equivalent to regressing against \emph{marginal} vector field $u_t$~\citep{tong2023improving}.
We can integrate Equation~\ref{eq:cfm} over time to obtain $p_\text{data}$ samples from noisy $x_0\sim q$.

\begin{figure*}[t]
\vskip -0.2in
\begin{center}
\includegraphics[width=\textwidth]{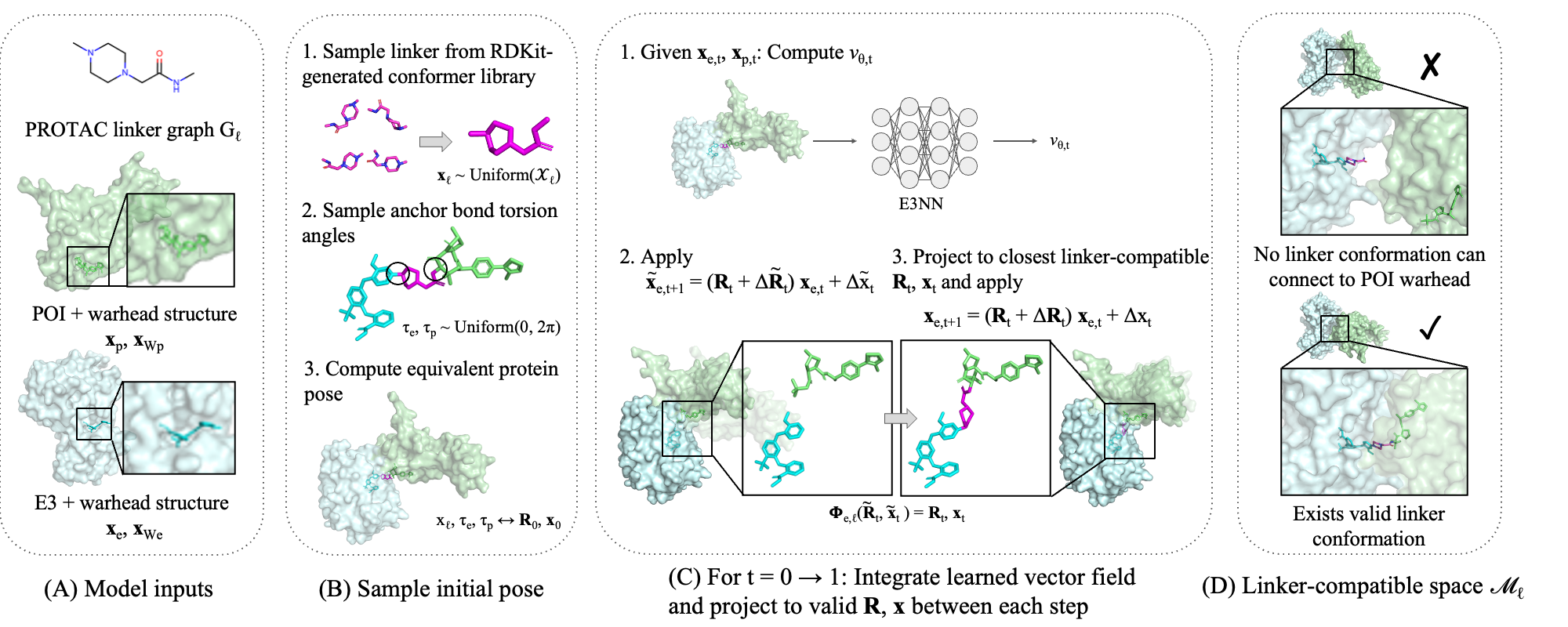}
\end{center}
\vskip -0.1in
\caption{
Overview of \ours{}, illustrated for PDB 7PI4.
Iteratively refine E3 ligase poses by 1) predicting a rotation/translation update from the learned vector field, 2) applying these vectors to approximate the next pose, and 3) projecting to the closest linker-compatible pose.
}
\label{fig:model_overview}
\vskip -0.2in
\end{figure*}

\section{Methods}

We present \ours{}, an iterative refinement framework based on flow matching for PROTAC-induced structure prediction.
As inputs, we are given unbound structures of an E3 ligase and protein of interest (POI), 
the molecular graph of the PROTAC, and (optionally) the E3 and POI warhead binding sites.
Our goal is to sample the PROTAC-induced complex of the E3 ligase and POI (Figure~\ref{motivation}).

Let $\xE\in\RR^{3\times\nE}, \xP\in\RR^{3\times\nP}$ denote the coordinates of the E3 ligase and POI.
Let $G = (\mathcal{V}, \mathcal{E})$ denote the molecular graph of the PROTAC.
$G$ is associated with subgraphs $G_{W_e}, G_{W_p}, G_\ell$, corresponding to the E3 warhead, POI warhead, and linker, where $G_\ell$ shares one edge (``anchor bond'') with each of the warheads.
We assume that the relative coordinates between warheads $\xWe, \xWp$ and the corresponding binding protein $\xE, \xP$ are given, either from ground truth or as the output of a protein-small molecule docking algorithm.\footnote{
The former is a common assumption throughout the PROTAC docking literature~\citep{zaidman2020prosettac}, while the latter is empirically an easy task (Appendix~\ref{subsec:warhead-docking}), as warheads are generally known binders.}
Our goal is to learn a generative model over
$P(\xE, \xlinker \mid \xP, G_\ell)$,
where we arbitrarily fix the POI and transform the E3.

In practice, we approximate $P(\xE, \xlinker \mid \xP, G_\ell)$
by sampling from $P(\xE\mid\xlinker, \xP, G_\ell)$ and optimizing $P(\xlinker\mid\xE, \xP, G_\ell)$.
We learn the former using an iterative refinement model, introduced in the next section, and we compute the latter using a deterministic search algorithm 
over a large set of generated linker conformations (Appendix~\ref{sec:linker_matching}).

\subsection{PROflow framework}\label{sec:CFM}

On a high level, our iterative refinement framework is as follows.
For more details, see Appendix~\ref{sec:model-appendix}.
\begin{enumerate}
    \item We define a linker-compatible space $\mathcal{M}_\ell$, which is the space of E3 ligase transformations that can be ``reached'' by any conformation of the linker under consideration (Figure~\ref{fig:model_overview}D).
    \item We sample from our prior over $\mathcal{M}_\ell$
    by sampling valid linker conformations $\xlinker$, anchor bond torsion angles, and converting to equivalent protein transformations (Figure~\ref{fig:model_overview}B).
    This ensures that all our samples remain within the linker-compatible space.
    \item We define our conditional flow map using the objects in SE(3) flow matching~\citep{yim2023fast}, followed by a projection to $\mathcal{M}_\ell$ (Figure~\ref{fig:model_overview}C).
    \item Finally, we optimize an objective that compares the predicted transformation, to that yielded by the analytical conditional vector field in SE(3) flow matching.
\end{enumerate}


Our learned vector field $v_\theta$ takes as input $G_\ell, \xE, \xP$, and time $t$.
The model predicts as output $\hat{x}_{\theta,t}, \hat{R}_{\theta,t}$ in the tangent space to $\text{SO}(3)\times\RR^3$.
We parametrize $v_\theta$ using E3NN, an equivariant graph convolution network~\citep{geiger2022e3nn}.
We process our inputs into protein interface point clouds, and include both geometric and chemical features.
The final prediction aggregates all vertices into a virtual node and applies an additional tensor product filter.

\subsection{Pseudo-data generation}\label{sec:data}

So far, we have assumed that our model has access to triplets $(\xE, \xP, \xlinker)$ for training.
In reality, there are fewer than twenty ternary PROTAC structures in the PDB, so it is infeasible to train on existing data.
Instead, we create a new pseudo-ternary dataset from binary protein-protein structures (DIPS)~\citep{townshend2019end}
and known linker graphs (PROTAC-DB)~\citep{weng2023protac}.

First, for each of the 1507 unique linkers $G_\ell$ reported in~\cite{weng2023protac}, we generate a set of 1024 conformations $\mathcal{X}_\ell$ using RDKit~\footnote{\url{https://www.rdkit.org/}} (linker library).
Next, for each binary protein-protein pair, we identify putative pocket candidates based on surface patches with high curvature.
During training, we sample one pocket per protein, per pair, from the top 4 pockets with the highest curvature on each protein.
Finally, for each pair of pockets, we identify an appropriate linker that has low RMSD to ``anchor bonds,'' placed in the center
of each pocket. 
See Appendix~\ref{sec:data-gen} for more details.
It's important to note that the training dataset and the ternary structures in our test set do not overlap.
\section{Experiments}

\begin{table*}[t]
\begin{small}\begin{center}
\vskip -0.1in
\setlength\tabcolsep{3 pt}
\caption{\textbf{Left:} re-docking, holo structures. \textbf{Right:} realistic docking, apo structures. \textbf{Format:} RMSD (\%$<$ threshold, 20\textup{\AA} for cRMSD, 10\textup{\AA} for iRMSD).
Methods marked $^*$ did not model the PROTAC. Runtimes marked $\dagger$ were performed exclusively on CPU (method does not support GPU). None of the methods utilize re-ranking or clustering for prediction selection.}
\label{tab:docking}
\begin{minipage}{.625\textwidth}
\begin{tabular}{l rr rr r r }
\toprule
& \multicolumn{6}{c}{Holo structures}
\\
Model & \multicolumn{2}{c}{cRMSD (\textup{\AA})}
& \multicolumn{2}{c}{iRMSD (\textup{\AA})}
& \multicolumn{1}{c}{Fnat}
& Time(s)
\\
\cmidrule(l{\tabcolsep}){1-1}
\cmidrule(l{\tabcolsep}){2-3}
\cmidrule(l{\tabcolsep}){4-5}
\cmidrule(l{\tabcolsep}){6-6}
\cmidrule(l{\tabcolsep}){7-7}

AF-Multimer$^*$ 
& 14.83 & (0.812) & 10.49 & (0.539) & 0.221 & 735$\dagger$ 
\\
DiffDock-PP$^*$ 
& 19.67 & (0.510) & 10.04 & (0.509) & 0.229
& 150 
\\
PROTACability 
& \textbf{12.45} &  (0.926) 
& 11.21 & (0.496) & 0.144 & 144$\dagger$ 
\\
PRossetaC 
& 13.06 & (0.954) & 10.03 & (0.487) &0.114 & 3015$\dagger$ 
\\ \midrule
PROflow 
& 12.85 & (\textbf{0.975}) & \textbf{8.20} & (\textbf{0.740}) & \textbf{0.264} & \textbf{44} 
\\
\bottomrule
\end{tabular}
\end{minipage}\begin{minipage}{.375\textwidth}
\begin{tabular}{rr rr r}
\toprule
\multicolumn{5}{c}{Apo structures}
\\
\multicolumn{2}{c}{cRMSD (\textup{\AA})}
& \multicolumn{2}{c}{iRMSD (\textup{\AA})}
& \multicolumn{1}{c}{Fnat}
\\
\cmidrule(l{\tabcolsep}){2-3}
\cmidrule(l{\tabcolsep}){4-5}
12.94 & (0.865) & 9.80 & (0.563) & 0.208 
\\
21.45 & (0.380) & 12.65 & (0.446) &0.206 
\\
27.58 & (0.641)
& 25.67 & 0.275 & 0.118 \\
15.33 & (0.856) & 12.13 & (0.367) &0.095  \\ \midrule
\textbf{12.16}
& (\textbf{0.995}) & \textbf{8.35} &  (\textbf{0.643}) & \textbf{0.236} \\
\bottomrule
\end{tabular}
\end{minipage}
\end{center}\end{small}
\vskip -0.1in
\end{table*}

\subsection{Docking experiments}
Our test set consists of 13 solved PROTAC ternary structures 
from the PDB, downloaded from PROTAC-DB~\citep{weng2023protac}.
We compare \ours{} with three classes of docking algorithms.
\begin{itemize}
    \item Unconditional protein-protein docking methods
    dock the E3 and POI without considering the PROTAC (Alphafold Multimer~\citep{evans2021protein}, DiffDock-PP~\citep{ketata2023diffdock}).
    \item Warhead only: PROTACability~\citep{pereira2023rational}
    assumes warhead binding sites are known for constrained protein-protein docking, but does not consider the linker.
    \item Full PROTAC: PRosettaC~\citep{zaidman2020prosettac} assumes warhead binding sites are known and uses precomputed linker lengths as distance constraints for protein-protein docking.
\end{itemize}
PROTACability and PRossetaC employ an additional clustering step tuned on part of our test set, so for fair comparison, we omit this in the main results.
See Table~\ref{table:prosettac-cluster} for results after post-processing.

Since PROTAC ternary structure prediction
focuses primarily on docking the E3 and POI,
we evaluate all methods based on common
protein-protein docking metrics.

\begin{itemize}
    \item Complex RMSD is determined by superimposing
    the predicted complex onto the ground truth complex
    using the Kabsch algorithm~\citep{kabsch1976solution}
    and computing the RMSD between the $\alpha$-carbons.
    \item Interface RMSD aligns only atoms in the
    ground truth binding interface (within 10\textup{\AA}
    of the binding partner) and computes the RMSD
    between their $\alpha$-carbons.
    \item Fnat is the fraction of native interface contacts
    (within 15\textup{\AA} of the binding partner)
    preserved in the predicted complex's interface.
\end{itemize}
These thresholds are slightly higher than
those in previous work~\citep{ganea2021independent}
since the distances between PROTAC complexes
are longer than between natural protein binders (Figure~\ref{Fig:ppi-distance-stats}).

\paragraph{Re-docking with holo structures}

We assess all models on their ability to recapitulate the E3-POI complex when provided their individual bound structures, as extracted from ternary structures (Table~\ref{tab:docking}).
\ours{} outperforms all baselines in interface RMSD and Fnat, and achieves similar performance to the best baseline in complex RMSD.
Of the baselines, those that do not consider the PROTAC at all (Alphafold-Multimer, DiffDock-PP) performed the worst in complex RMSD.
We provide visual examples of our predictions in Figure~\ref{fig:vis}.
All models were given access to 10 CPU cores and 1 NVIDIA A6000 GPU for the runtime analysis.
\ours{} achieves the fastest runtime of all models, with an average of 44 seconds per complex.

\paragraph{Docking with apo structures}

Table~\ref{tab:docking} evaluates whether models are able to produce the POI-E3 complex, given only their unbound structures (apo-equivalents from the PDB) collected by~\citet{pereira2023rational}.
These sequences are nearly identical to their holo counterparts, with 95\%+ sequence similarity.
For evaluation, we aligned apo-equivalents to the holo structures and placed the warheads into the corresponding pockets.
\ours{} was able to maintain high performance, while all other baselines (that take structural inputs) experienced a notable drop. 
This result indicates their high reliance on holo conformations, whose availability is an unrealistic expectation, as very few E3s have been co-crystallized with non-native protein targets.
The performance of AF-Multimer is improved mainly because of the alignment operation in apo structure docking procedure.


\subsection{PROTAC degradation activity experiments}

\begin{figure}[t]
\centering
\includegraphics[width=0.42\columnwidth]{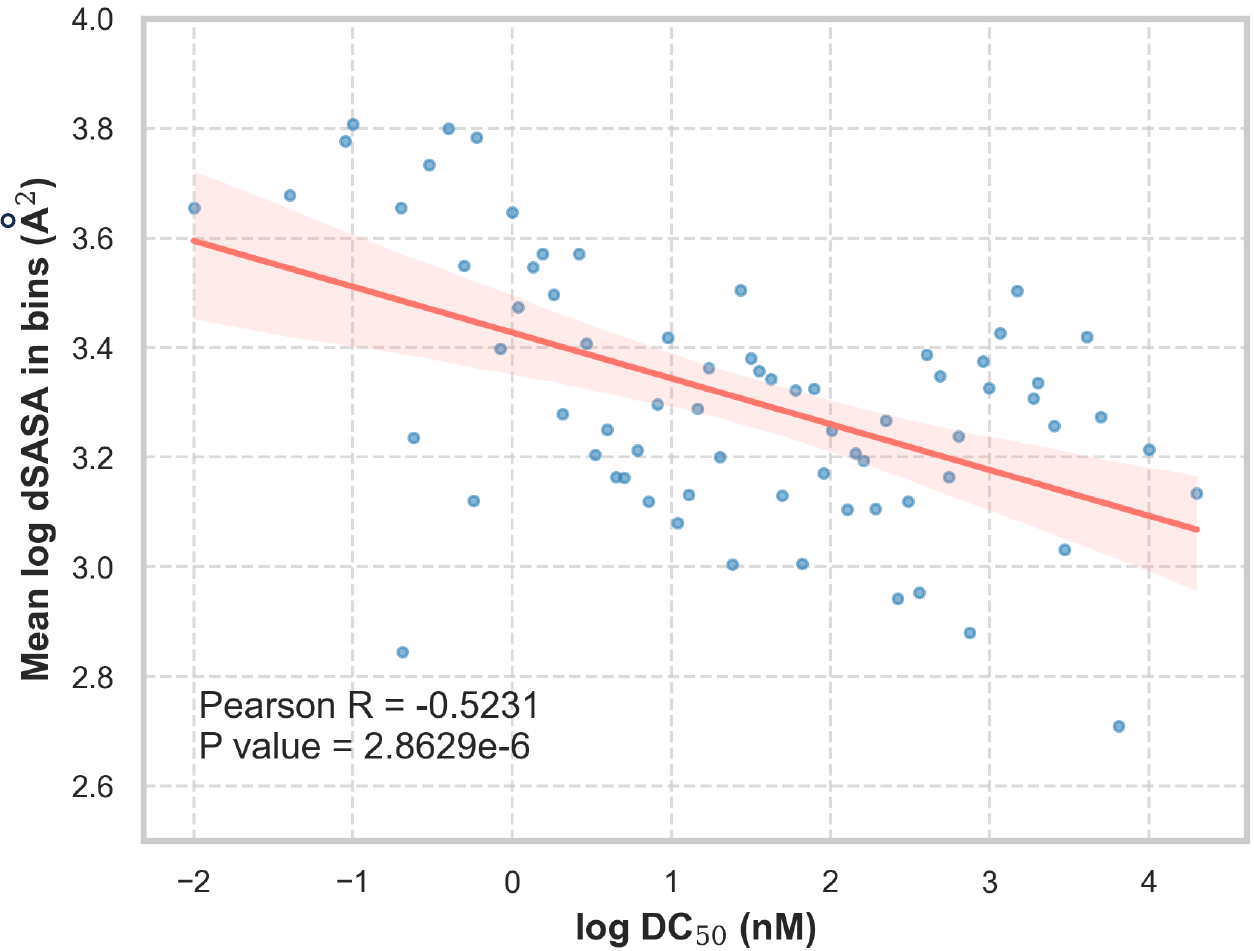}
\includegraphics[width=0.42\columnwidth]{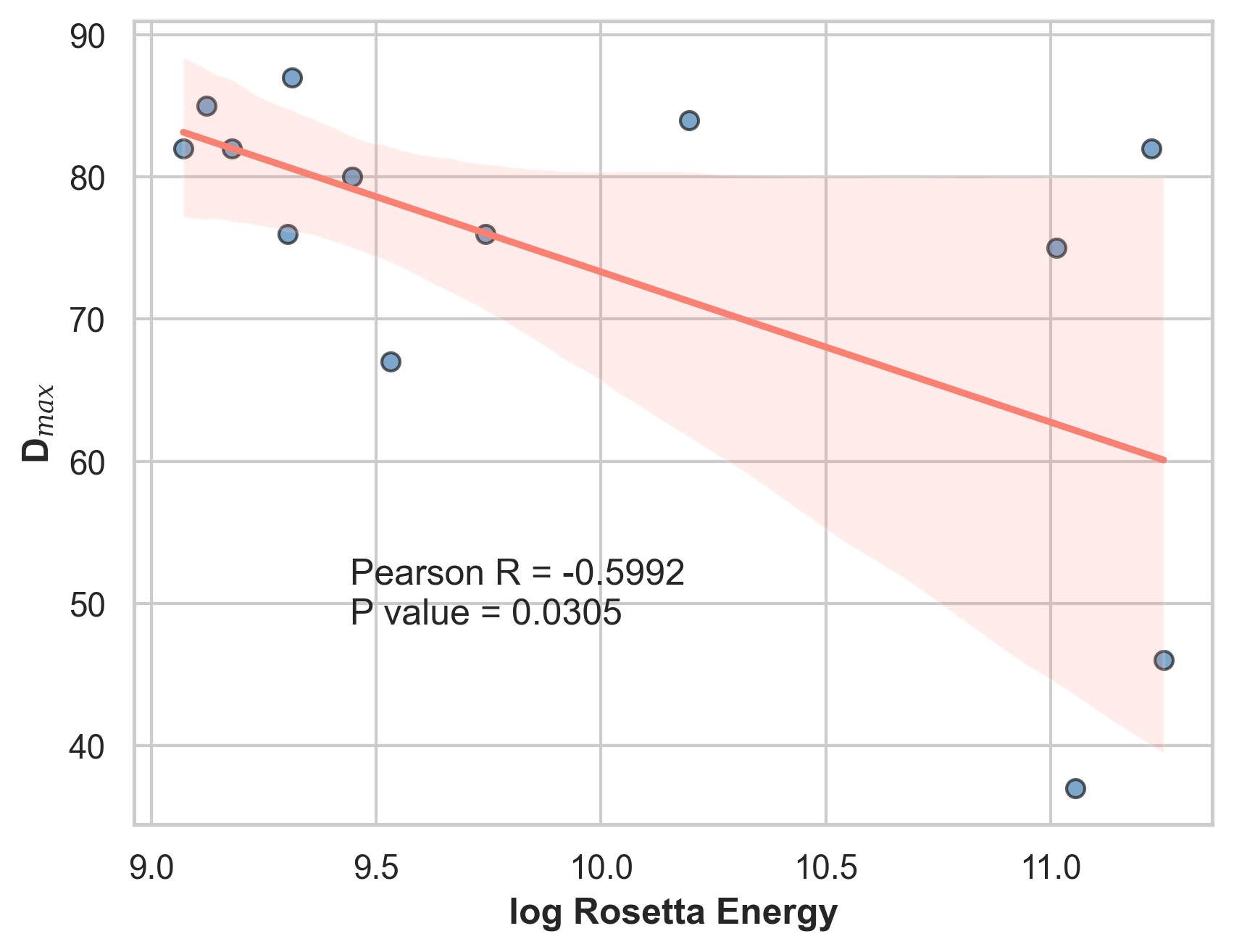}

\vskip -0.1in
\caption{
Left: DC$_{\text{50}}$ Solvent-accessible area buried at the interface(dSASA) of \ours{} predictions, over PROTAC-DB.
Right: D$_{\text{max}}$ Rosetta energy of \ours{} predictions, over PROTACs that vary only in linker.
}\label{fig:correlations}
\vskip -0.2in
\end{figure}

A major goal of predicting PROTAC-induced ternary structures is to facilitate their rational design.
Given that our model is able to predict high-quality structures at a fraction of the computational cost of all alternatives, while considering the entire PROTAC,
we are able to predict and assess PROTAC structures at scale.

\paragraph{Large scale PROTAC-DB study}

We collected 483 examples with annotated DC$_{\text{50}}$ values (the concentration of PROTAC required to degrade 50\% of the POI)~\citep{weng2023protac}, whose unbound E3 and POI structures were available in the PDB.
Nearly 70\% of the PROTACs under consideration (334)
share the same POI and E3 ligase with 4+ other PROTACs, so it is essential to model the linker in this analysis.
Of our baselines, only PRosettaC~\citep{zaidman2020prosettac} models the PROTAC linker.
However, it was computationally intractable to benchmark PRosettaC at scale (nearly one hour per structure).
We docked all warheads using Glide~\citep{friesner2004glide}.
We sampled 20 \ours{} structures per example and computed the 
solvent-accessible area buried at the interface (dSASA) using Rosetta~\citep{leaver2011rosetta3}.
We observed a statistically significant correlation between $\log$ dSASA and $\log \text{DC}_{\text{50}}$ (p-value $<0.001$) with a Pearson correlation coefficient of -0.5231 (Figure~\ref{fig:correlations}, left).
The direction of this correlation
is biologically intuitive:
PROTACs that induce larger interfaces have lower DC$_{\text{50}}$ and are more potent degraders.

\paragraph{VHL-SMARC2 case study}

In addition to DC$_{\text{50}}$, we also evaluate the relationship between the Rosetta energy of predicted PROTAC-induced E3-POI complexes and D$_{\text{max}}$ (maximum level of observed degradation).
Data for 13 D$_{\text{max}}$ values were sourced from the study by \citet{kofink2022selective}.
These PROTACs share identical warhead structures that target the same POI and E3 ligase and differ only in their linkers.
Figure~\ref{fig:correlations} (right) illustrates a statistically significant correlation (p-value $<0.05$), Pearson coefficient -0.5992) between log Rosetta energy and D$_{\text{max}}$.
Lower protein-protein energies signify a stronger POI and E3 ligase interaction,
so \ours{} predicted structures are able to filter out PROTAC designs with high degradation capacity.

 \section{Conclusion}

 In this work, we introduced \ours{}, an iterative refinement model for PROTAC ternary structure docking.
 To train this model, we generated a new pseudo-ternary dataset from binary 3D protein-protein data and 2D linker graphs.
 Empirically, \ours{} significantly outperformed baselines in interface RMSD and remained robust even when provided unbound structures.
 An order of magnitude faster than the best alternatives, \ours{} enabled the screening of large PROTAC libraries,
 revealing correlations between computed properties and degradation activity.
 In future work, we hope to explore deep learning models for other modalities of targeted degradation, including molecular glues.

\section*{Acknowledgements}

We thank Hannes Stark, Jason Yim, Bowen Jing, and Gabriele Corso for helpful discussions.

This material is based upon work supported by the National Science Foundation Graduate Research Fellowship under Grant No. 1745302. We would like to acknowledge support from the NSF Expeditions grant (award 1918839: Collaborative Research: Understanding the World Through Code), Machine Learning for Pharmaceutical Discovery and Synthesis (MLPDS) consortium, the Abdul Latif Jameel Clinic for Machine Learning in Health, and the DTRA Discovery of Medical Countermeasures Against New and Emerging (DOMANE) threats program.

\newpage

\bibliography{iclr2024_conference}
\bibliographystyle{iclr2024_conference}

\newpage
\appendix
\section{Iterative refinement framework}
\label{sec:model-appendix}

\paragraph{PROTAC-linker compatible space}
It is non-trivial to include $G_\ell$ explicitly in the model. However, generating multiple conformations $\mathcal{X}_\ell$ based on $G_\ell$ is an easy task~\citep{jing2023torsional}. In addition, the linker primarily functions as a positional scaffold and is less involved in the physicochemical interactions specific to the elements. Therefore, we approach $G_\ell$ conditional modeling alternatively by defining the space of $\text{SE}(3)$ transformations of the E3 ligase in the following space that respect a set of chemically plausible conformations $\mathcal{X}_\ell$ of the linker (Figure~\ref{fig:model_overview}D). 

\begin{definition}[Linker-compatible space]\label{def:space}
Let
$\mathbf{x}_{[\cdot] - \ell}$
denote the structure of a protein-warhead complex connected to linker $\ell$ at its respective anchor bond,
and 
let $\mathbf{x}_{[\cdot]}[b]\in\RR^{3\times2}$ denote the coordinates that correspond to anchor bond $b$.

For a given $\xE, \xP$ and $G_\ell$,
a transformation $(R, x) \in \text{SE}(3)$ belongs to \emph{linker-compatible space}
$\mathcal{M}_\ell$
if there exists a conformation $\xlinker$
that connects the POI and E3 under $(R, x)$:
\begin{equation}
    \exists \xlinker\in \mathcal{X}_\ell,~
    \left(
    R~\mathbf{x}_{e-\ell} + x
    \right)[b_p]
    = \xP[b_p],
\end{equation}
where $b_p$ is the POI anchor bond.
\end{definition}

It is non-trivial to define the standard conditional flow matching objects within $\mathcal{M}_\ell$.
Instead of operating directly over this set,
we leverage the objects of SE(3) flow matching~\cite{yim2023fast} and project back into $\mathcal{M}_\ell$ at each step --- a procedure we call \emph{pseudo}-flow matching.

Similar to standard flow matching, we first define a prior distribution on $\mathcal{M}_\ell$.
Then, we define the \emph{pseudo}-conditional flow map from the prior to the target data distribution by interpolating on SE(3) and projecting the intermediate points from SE(3) to $\mathcal{M}_\ell$.
Finally, we discuss our modified training objective and touch upon symmetry considerations.

\paragraph{Prior distribution}\label{sec:prior_distribution}

To design a flow matching framework,
we require an easy-to-sample $q_\ell(z)$ over $\mathcal{M}_\ell$.
However, it is not necessarily easy to sample from this space:
selecting an arbitrary $(R,x)\in\text{SE}(3)$ does not ensure the existence of a compatible $\xlinker$.

Therefore, we take an inverted approach: instead of directly operating over the protein, we first sample a valid linker conformation, and then compute the equivalent protein transformation (Figure~\ref{fig:model_overview}B).
To summarize, sampling from $q_\ell$ can be decomposed into three steps.
\begin{enumerate}
    \item Sample linker conformation $\xlinker\sim\text{Uniform}(\mathcal{X}_\ell)$.
    \item Sample torsion angles for each anchor bond $\tau_e, \tau_p \sim \text{Uniform}(0, 2\pi)$.
    \item Map to equivalent E3 ligase transformation $(R, x)$.
\end{enumerate}
The closed-form solution for the mapping $\xlinker, \tau_e, \tau_p \mapsto (R, x)$ can be obtained by absorbing the anchor bond torsion angles $(\tau_e, \tau_p)$ and angles within the linker's local structure into rotation matrices. A detailed derivation is presented in Section~\ref{sec:q_sample}.

Intuitively, this procedure can be interpreted as sampling from the linker conformation energy-weighted distribution over $\mathcal{M}_\ell$. Transformations $(R, x)$ that correspond to stable linker conformations are sampled more frequently, and $(R, x)$ that result from different torsion angle settings of the same linker conformation are sampled equally often.

\paragraph{Pseudo-conditional flow map}

Interpolating along the geodesic paths in SO(3) and $\RR^3$ gives rise to the
following conditional flow maps~\cite{yim2023fast}:
\begin{align}
 \tilde{R_t} &= \exp_{R_0}(t\log_{R_0}(R_1))\label{eq:pt_r}\\
 \tilde{x_t} &= (1-t)x_0 + t \cdot x_1\label{eq:pt_x}.
\end{align}
However, this interpolation does not guarantee that we remain within $\mathcal{M}_\ell$.
Instead,
we introduce the following projection
to define a reasonable path within $\mathcal{M}_\ell$.

\begin{definition}[Projection to $\mathcal{M}_\ell$]\label{def:project}
Given $(R,x)\in\text{SE}(3)$,
we define $\Phi_{\ell,e}:\text{SE}(3)\to\mathcal{M}_\ell$
\begin{equation}\label{eq:phi}
    \Phi_{\ell,e}(R, x) = 
    \mathop{\arg\min}_{(R^*, x^*) \in \mathcal{M}_l} \text{RMSD}\left( (R \xE + x), (R^* \xE + x^*)\right)
\end{equation}
where $\text{RMSD}(\mathbf{x}, \mathbf{x}') = \sqrt{1/n \cdot \sum_i^n \|\mathbf{x}_i - \mathbf{x}'_i\|^2 }$.
\end{definition}
This procedure allows us
to compare different transformations based on their effects on $\xE$, and
to implicitly update the $\xlinker$ and $\tau_e,\tau_p$ under consideration at each step.
We implement this mapping via the
conformer matching optimization procedure used in~\citet{jing2023torsional} (Appendix~\ref{sec:linker_matching}).

We define the pseudo-conditional flow map over $\mathcal{M}_\ell$,
\begin{equation}
 (R_t, x_t) = \Phi_{\ell,e}(\tilde{R_t}, \tilde{x_t})\label{eq:target}
\end{equation}
where $\tilde{R_t}, \tilde{x_t}$ are given by Equations~(\ref{eq:pt_r}, \ref{eq:pt_x}).
At $t=0$, our linker-centric sampling procedure ensures $(R_0, x_0)\in\mathcal{M}_\ell$, and at $t=1$, $(R_1, x_1)$ are computed directly from data.

\paragraph{Training objective}

Due to the nature of the projection to $\mathcal{M}_\ell$, it is difficult to obtain an analytical formula for the vector field $u_t$.
If we first ignore the projection, standard SE(3) flow matching yields a closed form solution by differentiating Equations (\ref{eq:pt_r}, \ref{eq:pt_x}) with respect to $t$ ~\cite{yim2023fast}:
\begin{align}
    \dot{x}_{t} &\approx \frac{x_{1} - \tilde{x}_{t}}{1-t},
    &\dot{R}_{t} \approx \frac{\log _{\tilde{R}_{t}}\left(R_{1}\right)}{1-t}.
\end{align}
The simplest solution is to apply the standard conditional flow matching loss (Equation \ref{eq:loss_f}) to these terms.
However, this approach leads to unstable performance, due to the irregularity of $\mathcal{M}_\ell$ (analysis in Appendix~\ref{sec:stability}).

Therefore, we optimized an alternative,
which compares the effect of predicted transformations on the coordinates.
Let $z\sim q$ and $(R_t, x_t)\sim p_t(x\mid z)$, then:
\begin{align}
    \tilde{\mathbf{x}}_{e,t + \Delta t} &= \dot R_t \Delta t(R_t \mathbf{x}_{e,t})
    + (x_t + \dot x_t \Delta t)\\
    \hat{\mathbf{x}}_{e,t + \Delta t} &= \hat R_{\theta,t} \Delta t(R_t \mathbf{x}_{e,t})
    + (x_t + \hat x_{\theta,t} \Delta t)
\end{align}
where $\hat{x}_{\theta,t}, \hat{R}_{\theta,t}$ are the outputs of the learned vector field $v_\theta$,
and $\Delta t$ is set to a small constant.
We minimize
\begin{align}
    \mathcal{L}_{\mathcal{M}_l} = \mathbb{E}&_{t, q(z), p_t(x\mid z)}
    \text{RMSD}(\tilde{\mathbf{x}}_{e,t + \diff t},
    \hat{\mathbf{x}}_{e,t + \diff t}).
\end{align}

\paragraph{Sampling}

During inference, we integrate (\ref{eq:cfm}) from $t\in[0,1]$ using an Euler solver~\citep{chen2018neural}.
We apply $\Phi_{\ell,e}$ between each step to ensure that our transformations remain within $\mathcal{M}_\ell$ (Figure~\ref{fig:model_overview}C and Algorithm~\ref{alg:sample}).

\paragraph{SE(3) equivariance}

To ensure that our likelihood is SE(3)-equivariant, we require a SE(3) invariant prior and SE(3)-equivariant updates~\cite{xu2022geodiff}.
Our prior distribution is SE(3) invariant since the torsion angles of anchor bonds $\tau_e, \tau_p$ are SE(3) invariant,
we express $\xlinker$ relative to the input $\xP$ with local structures fixed.
The learned vector field
$v_\theta$ is parametrized by a SE(3) equivariant architecture based on~\citet{geiger2022e3nn}.

\subsection{Model implementation}
\label{sec:model-details-appendix}

\paragraph{Input representations} 

Given the two input protein structures, we first compute their surfaces using MSMS and triangulate them into point clouds of vertices (same as in Section ~\ref{sec:data}).
We featurize each vertex with geometric and chemical features~\citep{gainza2020deciphering, sverrisson2021fast}, including its coordinates, curvature, normal vector, 6 closest non-hydrogen atoms, and the vectors between these atoms and the vertex itself.

Prior work has shown that when protein-protein interfaces are known, removing non-interface residues from consideration increases both computational efficiency and docking performance~\citep{pereira2023rational}.
Since warhead poses are determined ahead of time, we follow the same approach and discard any vertices more than 20\textup{\AA} from the surface of the binding partner, at initialization.
Though this definition of interface depends on the initial $\xlinker$, this threshold preserves approximately one-third of most proteins, which is usually a superset of the actual interface.

\paragraph{SE(3)-equivariant architecture}

We implement $v_\theta$ using a 6-layered E3NN, a SE(3)-equivariant graph convolutional network~\citep{geiger2022e3nn}.
Within each protein point cloud, we draw edges between the 3 nearest neighbors of each vertex, and across the protein interface, we add edges between vertices within 15\textup{\AA} apart.
To produce the SE(3)-equivariant outputs $\hat{x}_{\theta,t}, \hat{R}_{\theta,t}$ from the final layer,
we aggregate information from all vertices into a virtual node, placed at the center of all E3 ligase $\alpha$-carbon atoms (including those outside the interface), and apply an additional tensor product filter. Additional information can be found in Appendix~\ref{sec:appd_arch}.

\paragraph{Steric guidance}

Though our E3NN focuses exclusively on the protein interface surface, we would like to avoid steric clashes from parts of the proteins outside of their interfaces.
We add an additional potential to the gradient
to bias the model away from transformations that would otherwise cause clashes.
We modify our vector field as follows,
\begin{equation}\label{eq:steric}
    u_{t}(x) \leftarrow u_{t}(x) + \sigma_t
    \nabla_x\sum_{i=1}^{m_p}\sum_{j=1}^{m_e}[-\max(0, d_{ij} - \rho)],
\end{equation}
where $\sigma_t = \sigma_{\min}^{1-t} \cdot \sigma_{\max}^{t}$ is an exponential scheduler,
$d_{ij}$ is the Euclidean distance between $\alpha$-carbons of residues $i,j$, and $\rho=$ 2\textup{\AA} is the contract threshold.

\begin{figure*}[t]
\begin{center}
\includegraphics[width=0.98\textwidth]{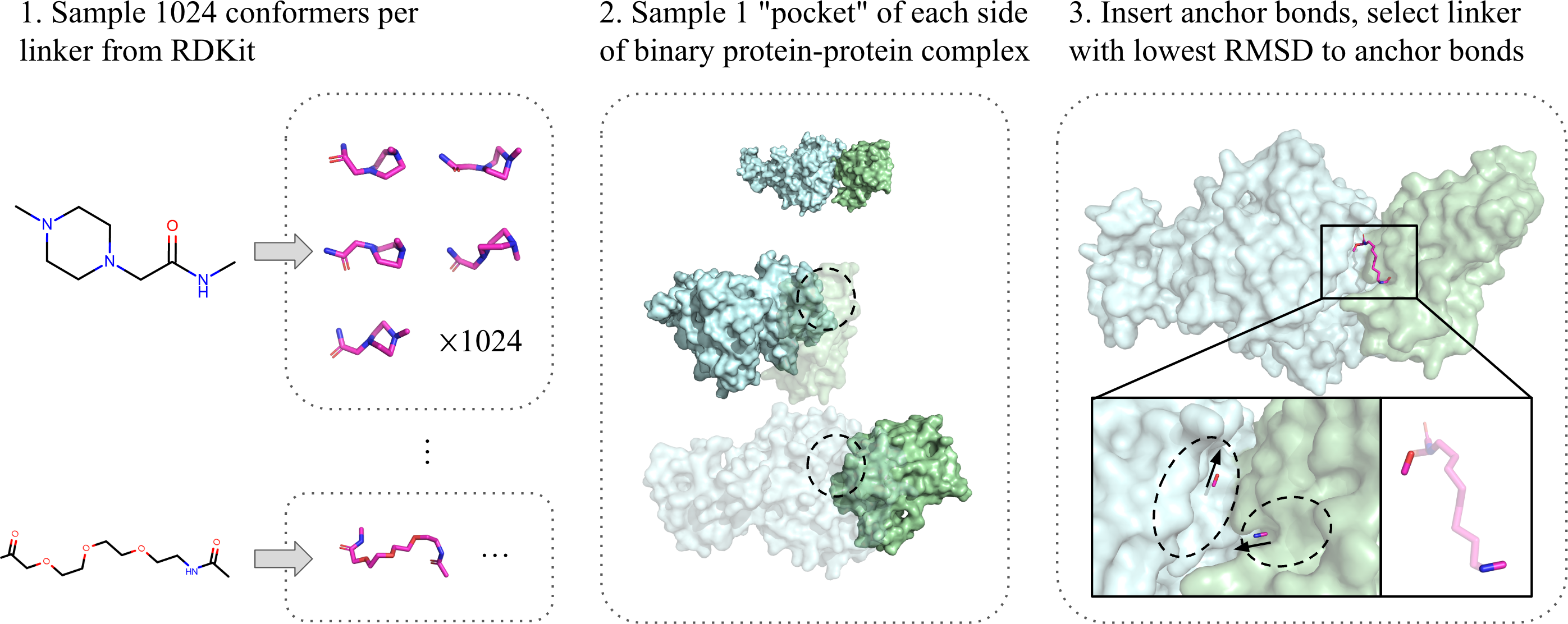}
\label{fig:data-gen}
\end{center}
\vskip -0.1in
\caption{Pseudo-data generation procedure. 1) Generate PROTAC linker conformation library. 2) Identify high-curvature putative pockets and sample 1 per protein in binary protein-protein complex. 3) Match each protein pair to linker with lowest RMSD from anchor bonds.}
\vskip -0.2in
\end{figure*}

\subsection{Data generation procedure}
\label{sec:data-gen}
\paragraph{Linker conformation library} For each of the 1507 unique linkers $G_\ell$ reported in~\cite{weng2023protac}, we generate a set of 1024 conformations $\mathcal{X}_\ell$ using RDKit~\footnote{\url{https://www.rdkit.org/}}.

\paragraph{Pseudo-binding sites} We identify putative warhead binding sites on each side of the binary protein-protein interface.
Since protein interfaces are too flat for standard pocket prediction methods~\cite{le2009fpocket},
we opt for a simpler geometric approach.
We compute each protein's surface using MSMS~\cite{sanner1996reduced} and triangulate with a vertex density of 0.5/\textup{\AA}$^2$.
Putative pockets are obtained by clustering the top 200 vertices with highest curvature, based on pairwise Euclidean distance.
Pseudo-anchor bonds are inserted at each cluster's center of mass, pointing towards the average of the vertices' normal vectors.
For each protein, we extract the top 4 highest curvature clusters as pockets. During training, we sample one pocket per protein uniformly at random, from a total of 16 pairs.

\paragraph{Linker matching} For each pair of protein pockets, we select the best-matched linker from our conformation library $\{\mathcal{X}_\ell\}$
based on Kabsch alignment of the anchor bonds~\cite{kabsch1976solution}.
While this step only considers the two endpoints of each linker, most PROTAC linkers tend to be simple chains, which rarely result in any clashes with the protein complexes.

In total, this process allowed us to create approximately 685k triplets (42.8k protein pairs in DIPS $\times$ 16 pocket combinations per protein pair).
Our model was trained on these pseudo-data from DIPS and finetuned on a similarly created dataset of binary E3 ligase structures (Appendix~\ref{sec:e3_data}).

\section{Mathematical Derivations}
\subsection{Solution for equivalent prior distribution sampling}~\label{sec:q_sample}
In Section~\ref{sec:prior_distribution}, we decompose the sampling from prior distribution into sampling linker conformation and torsion angles. Here, we introduced how to derive the equivalent transformation $(R, x)$ from linker conformation $\xlinker$, $\tau_e$ and $\tau_p$. 
At first, we reduce the unnecessary degrees of freedom in $\xlinker$ by representing it as the $m$ torsion angles within the linker rotatable bonds $b_r$ as $\tau_\ell \in SO(2)^m$.

Because linkers usually adopt linear molecular graphs, the rotation matrices could be used for updating the ternary structure in sequential order. We reorder the atoms in the linker according to the distance to the POI side anchor bond to adopt the following permutation:
\begin{equation}
    \forall i, j: i < j \Rightarrow D_{\mathcal{G}_\ell}(i, 0) < D_{\mathcal{G}_\ell}(j, 0)
\end{equation}
$D_{\mathcal{G}}(i, j)$ is defined as the shortest path length between node $i$ and node $j$ in graph $\mathcal{G}$. The anchor bond from the POI side is denoted as edge $(0,1)$ in the $\mathcal{G}_\ell$. In addition, we also reordered the rotatable bonds as:
\begin{align}
    b_{r_i} &= (i_1, i_2), \\
    b_{r_j} &= (j_1, j_2) \\
    \forall i, j &: i < j \Rightarrow (i_1 < j_1 ~\text{and}~ i_2 < j_2)
\end{align}
Then, we transform all angles into rotation matrices:
\begin{align}
&R(\mathbf{n}, \alpha) = \begin{bmatrix}
    \mathbf{n}^2_x + (1-\mathbf{n}^2_x)\cos \alpha & \mathbf{n}_x \mathbf{n}_y (1 - \cos \alpha) - \mathbf{n}_z \sin \alpha & \mathbf{n}_x \mathbf{n}_z (1 - \cos \alpha) + \mathbf{n}_y \sin \alpha \\
    \mathbf{n}_x \mathbf{n}_y (1 - \cos \alpha) + \mathbf{n}_z \sin \alpha & \mathbf{n}^2_y + (1-\mathbf{n}^2_y)\cos \alpha & \mathbf{n}_y \mathbf{n}_z (1 - \cos \alpha) - \mathbf{n}_x \sin \alpha \\
    \mathbf{n}_x \mathbf{n}_z (1 - \cos \alpha) - \mathbf{n}_y \sin \alpha & \mathbf{n}_y \mathbf{n}_z (1 - \cos \alpha) + \mathbf{n}_x \sin \alpha & \mathbf{n}^2_z + (1-\mathbf{n}^2_z)\cos \alpha
\end{bmatrix}\\
&R_{\ell_i} = R(\frac{\xlinker[b_{r_i}]}{\|\xlinker[b_{r_i}]\|}, \tau_{\ell_i}), R_{\tau_e} = R(\frac{\mathbf{x}_{W_{e}}[b]}{\|\mathbf{x}_{W_{e}}[b]\|}, \tau_e), R_{\tau_p} = R(\frac{-\mathbf{x}_{W_{p}}[b]}{\|\mathbf{x}_{W_{p}}[b]\|}, \tau_e)
\end{align}
The transformation of rotating a substructure $\mathbf{x}[\cdot]$ around the rotation axis $\mathbf{x}[i,j]$ with angle $\tau$ can be written as:
\begin{equation}
    Trans(\tau, \mathbf{x}[i,j]) \mathbf{x}[\cdot] = R(\frac{\mathbf{x}[i,j]}{\|\mathbf{x}[i,j]\|}, \tau)(\mathbf{x}[\cdot] - \mathbf{x}[i]) + R(\frac{\mathbf{x}[i,j]}{\|\mathbf{x}[i,j]\|}, \tau)\mathbf{x}[i] 
\end{equation}
We denote given initial structure as $\mathbf{x}_{e_{int}}$. As a result, mapping the sampled linker conformation $\xlinker$ and torsion angles ($\tau_e$, $\tau_p$) into the E3 ligase relative position $\mathbf{X}_e$ formalizes a closed function as:
\begin{align}
    \mathbf{x}_e =  [Trans(\tau_e, \mathbf{x}_{W_{e}}[b]) \cdot Trans(\tau_{\ell_m}, \xlinker[b_{r_m}]) \cdots Trans(\tau_{\ell_0}, \xlinker[b_{r_0}]) \cdot Trans(\tau_p, \mathbf{x}_{W_{p}}[b])] \mathbf{x}_{e_{int}}
\end{align}

\section{Implementation Details}
\subsection{Search algorithm for linker matching}~\label{sec:linker_matching}
In this section, we introduce the algorithm applied for $P(\xlinker \mid \xE, \xP, G_\ell)$ and $\Phi_{\ell, e}$. Though these 2 processes are utilized in different steps of the generation, a conditional conformation generation can be implemented to solve both. According to the definition of anchor bonds, their coordinates $\xlinker[b]$ are determined when $\mathbf{x}_{W}[b]$ is determined by $\xE$ and $\xP$ because we keep the warhead-protein complex structure fixed. 

Both $P(\xlinker \mid \xE, \xP, G_\ell)$ and $\Phi_{\ell, e}$ can be reduced to searching process for finding $\xlinker \in \mathcal{X}_\ell$ to minimize the RMSD objective in Equation~\ref{eq:phi}. The optimum for this function is guaranteed to be 0 for $P(\xlinker \mid \xE, \xP, G_\ell)$, because $\xE$ lies in the constraint subspace, $\mathcal{M}_\ell$. However, for $\Phi_{\ell, e}$, the optimum is not guaranteed to be 0. As a result, the position of E3 ligase is updated in $\Phi_{\ell, e}$ to ensure $\text{RMSD}(\xlinker[b], \mathbf{x}_{W}[b]) = 0$ by aligning $\mathbf{x}_{W_e}[b]$ with $\xlinker[b_e]$.

Therefore, to minimize the objective, $\text{RMSD}\left( (R \xE + x), (R^* \xE + x^*)\right), \text{s.t.}~(R^*, x^*) \in \mathcal{M}_\ell$ we apply a scoring function with a optimization algorithm. To score a given linker structure $\xlinker^*$, we first align the $\xlinker^*[b]$ with $\mathbf{x}_{W}[b]$ derived from $\xE$ and $\xP$. We add interpolation coordinates within $\xlinker^*[b_p]$ and $\mathbf{x}_{W}[b_p]$ to strengthen the weighting in the Kabsch alignment, as we want to achieve precise alignment of $\xlinker^*[b_p]$ and $\mathbf{x}_{W}[b_p]$. This step is pivotal for evaluation on $\xlinker^*[b_e]$ and $\mathbf{x}_{W}[b_e]$ from the E3 ligase side, since our RMSD is defined for E3 ligase coordinates $\xE$. We denote the function of calculating the centroid coordinate as $c$, then $R^*, x^*$ is written as:
\begin{align}
R^* &= R(\frac{\overrightarrow{\xlinker^*[b_e]} \times \overrightarrow{\mathbf{x}_{W}[b_e]}}{\|\overrightarrow{\xlinker^*[b_e]}\| ~\|\overrightarrow{{\mathbf{x}_{W}[b_e]}}\|}, \arccos (\frac{\overrightarrow{\xlinker^*[b_e]} \cdot \overrightarrow{\mathbf{x}_{W}[b_e]}}{\|\overrightarrow{\xlinker^*[b_e]}\| ~\|\overrightarrow{{\mathbf{x}_{W}[b_e]}}\|}))\\
x^* &= R^*(c(\xE) - c(\mathbf{x}_{W}[b_e])) - (c(\xE) - c(\mathbf{x}_{W}[b_e])) + (c(\xlinker^*[b_e]) - c(\mathbf{x}_{W}[b_e]))
\end{align}
At each scoring step, we calculate the above $(R^*, x^*)$ to the RMSD evaluation in Equation~\ref{eq:phi}. We represent $\xlinker$ as the torsion angles $\tau_\ell$ and optimize the angles using an optimization algorithm given the above scoring function. To balance between accuracy and efficiency, the DIRECT algorithm is applied for $\Phi_l$ when we solve the ODE for sampling, and the evolution algorithm is applied for $P(\xlinker \mid \xE, \xP, G_\ell)$.
 
\subsection{Procedure for protein-protein training dataset curation}~\label{sec:e3_data}
In this section, we introduce the procedure for preparing the binary protein-protein data for the data generation pipeline (Section~\ref{sec:data}). We mainly train our data on the Database of Interacting Protein Structures (DIPS)~\citep{townshend2019end} dataset and finetune it on the E3 ligase dataset.

For the DIPS dataset, we simply run the script from the official repository~\footnote{https://github.com/drorlab/DIPS}. Unlike protein-protein docking researches that assign the larger protein chain as receptor and the shorter as ligand, we randomly assign (receptor, ligand) to the chains to increase the size of training data and because of the fact that E3 ligase is not always smaller than the POI. Though there are already 42,826 data points within the DIPS dataset, we still want to curate a dataset that only focuses on protein-protein complexes of E3 ligase + X, where X could be any protein chain. Firstly, we retrieve all the protein structures with multiple chains from PDB by using the E3 ligase Uniprot ID from PROTAC-DB~\citep{weng2021protac}. Then we screened these PDB files by extracting the E3 ligase chain and keeping the surrounding protein chain if it possesses more than 40 residues and more than 5 residues are close to the E3 ligase chain by the threshold of 8\textup{\AA}. By extracting the protein chain files, we curated a protein-protein dataset of 883 E3-ligase+X data. In the training, the E3 ligase is treated as the ligand and the binding chain is assigned as the pseudo-POI.

\subsection{SE(3) model architecture details}~\label{sec:appd_arch}
As outlined in Section~\ref{sec:model-appendix}, our parametrized vector field is implemented using a SE(3) convolutional network, following the architecture of the score model in Diffdock~\citep{corso2022diffdock}. In the following sections, we denote $\otimes_W$ as the spherical tensor product with weights $W$ and $\oplus$ as the normal vector addition. The architecture is decomposed into three main layers: the embedding layer, interaction layer, and output layer, as detailed below.

\paragraph{Embedding layer} Though previous protein-protein docking works~\citep{ganea2021independent, ketata2023diffdock} prefer to use residue-types and the coordinates of $\alpha$-carbons to represent rigid protein structures, PROTAC docking requires fine-grained representations as the protein-protein interface are predetermined by the warhead positions. In our work, we represent protein as surface representations. At first, we process the chemical features of the nearest atom into the atom embedding and vertex representations (curvature, size, type, normal vector) calculated by MSMS into the vertex embeddings with the same size of atom embedding. Both atom embedding and vertex embedding include scalar representations and vector representations. Then we aggregate 6 atom embeddings into the corresponding vertex embeddings by the following fully connected tensor product:
\begin{equation}
    \mathbf{h}_{v} = \frac{1}{6 + 1} \sum_{\mathbf{h}_{i} \in (\mathbf{h}_a, \mathbf{h}_v)} \sum_{\mathbf{h}_{j} \in (\mathbf{h}_a, \mathbf{h}_v)} \mathbf{h}_i \otimes_{W} \mathbf{h}_j
\end{equation}
In the embedding layer, $W$ are trainable parameters. In order to make \ours{} scalable to large protein-protein interfaces, we then cluster the vertices into a fixed number(256 in our model) of patches. The patch representations are calculated by averaging the vertex representations and place in the centroid of the corresponding vertices. 

\paragraph{Interaction layer} We defined the mechanism for updating the patch representations following the convolutional network in Diffdock~\citep{corso2022diffdock}. For a given patch with representation $\mathbf{h}_p$ that is connected to $\mathcal{N}_{p}$ other patches, for each message passing procedure, the representation is updated as:
\begin{align}
    \mathbf{h}_{p} \leftarrow \mathbf{h}_{p} {\oplus} \mathbf{B N}\left(\frac{1}{\left|\mathcal{N}_{a}\right|} \sum_{n \in \mathcal{N}_{a}} Y\left(r_{np}\right) \otimes_{\psi_{np}} \mathbf{h}_{n}\right), \psi_{np} = \Psi(\mathbf{h}_{p}, \mathbf{h}_{n}, \|r_{np}\|^2)
\end{align}
Here, $Y$ represents spherical harmonics, and BN stands for equivariant batch normalization. $\Psi$ denotes the trainable MLP responsible for computing the weights for the tensor product. In practice, we construct a k-nearest-neighbor graph (k=3) for message passing within patches for the same protein and a radius graph with a cutoff of 15\textup{\AA} for message passing between patches belonging to the POI and E3 ligase, respectively. In each interaction layer, the representations are initially updated by aggregating information from the same protein and subsequently exchanging information across proteins.

\paragraph{Output layer} It was crucial for us to aggregate the information back to the coordinate system of the full protein from representations from protein surfaces because the training objective is defined on full protein $\alpha$-carbon coordinates. An additional virtual node is placed at the centroid of E3 ligase $\alpha$-carbons. For all the patches of E3 ligase $\mathcal{V}_e$ and the centroid virtual node $v_c$, the message is passed to $v_c$ as:
\begin{equation}
    \mathbf{v_c} \leftarrow \frac{1}{\left|\mathcal{V}_{e}\right|} \sum_{n \in \mathcal{V}_{e}} Y\left(\hat{r}_{c n}\right) \otimes_{\psi_{c n}} \mathbf{h}_{n}, 
    \psi_{c n} = \Psi(\mathbf{h}_{n}, \|r_{c n}\|^2)
\end{equation}
To parametrize the vector field in the tangent space of SE(3), $\mathbf{v_c}$ is a single odd and a single even vector. We sum the odd representation with the even representation as the final prediction. For loss computation in training, we update the E3 ligase coordinates with our prediction and the ground truth score. The RMSD between the score updated coordinate and the predicted score updated coordinate is our training objective.

\renewcommand{\algorithmicrequire}{\textbf{Input:}}
\renewcommand{\algorithmicensure}{\textbf{Output:}}
\begin{algorithm}
\caption{Sampling algorithm of \ours{}}\label{alg:sample}
\begin{algorithmic}[1]
\REQUIRE{Initial $\mathbf{x}_p$ and $G_\ell$}
\ENSURE{Bound $\mathbf{x}_e$, $\mathbf{x}_{\ell}$}
\STATE{Sample from prior distribution $p_0$: $\xlinker\sim\text{Uniform}(\mathcal{X}_\ell)$, $\tau_e, \tau_p \sim \text{Uniform}(0, 2\pi)$}
\STATE{Map to equivalent $x_0 = (R_0, X_0)$}
\FOR{t in 0, $\Delta t, 2 \Delta t, \cdots 1$}
    \STATE{Compute $\hat{R}_t, \hat{X}_t$ from learned vector field $v_{\theta}$}
    \STATE{Update and refine to $\mathcal{M}_\ell$: $X_t \rightarrow \Phi_{\ell,e}(\hat{R}_t \Delta t x_t + \hat{X}_t \Delta t)$}
    \STATE{Compute the gradient of rotation and translation of $u(t)(X_t)$ in Eq.\ref{eq:steric}: $\hat{R}_{st}$,$ \hat{X}_{st}$}
    \STATE{Update and refine to $\mathcal{M}_\ell$: $X_t \rightarrow \Phi_{\ell,e}(\hat{R}_{st} \Delta t x_t + \hat{X}_{st} \Delta t)$}
\ENDFOR
\STATE{$\mathbf{x}_e = x_1$ and search $\mathbf{x}_{\ell}$ using the method described in Sec.~\ref{sec:linker_matching}}
\end{algorithmic}
\end{algorithm}

\section{Additional Experimental Results}
\label{sec:additional-results}

\subsection{Visual examples}

\begin{figure*}[h]
\begin{center}
\includegraphics[width=0.95\textwidth]{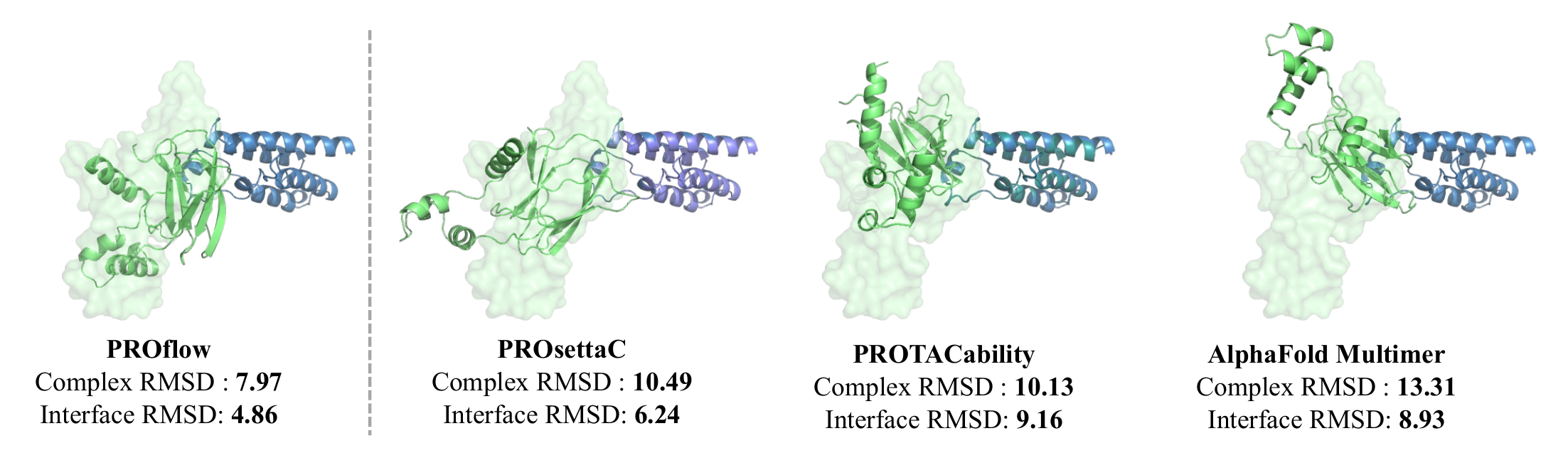}
\end{center}
\vskip -0.1in
\caption{Visualized results of structures (PDB 6HAX) predicted by different methods. The ground truth E3 ligase pose is represented as a semitransparent surfaces.}
\label{fig:vis}
\end{figure*}

\subsection{Protein-protein interface distance distribution}
\begin{figure}[h]
\centering
\includegraphics[width=0.6\columnwidth]{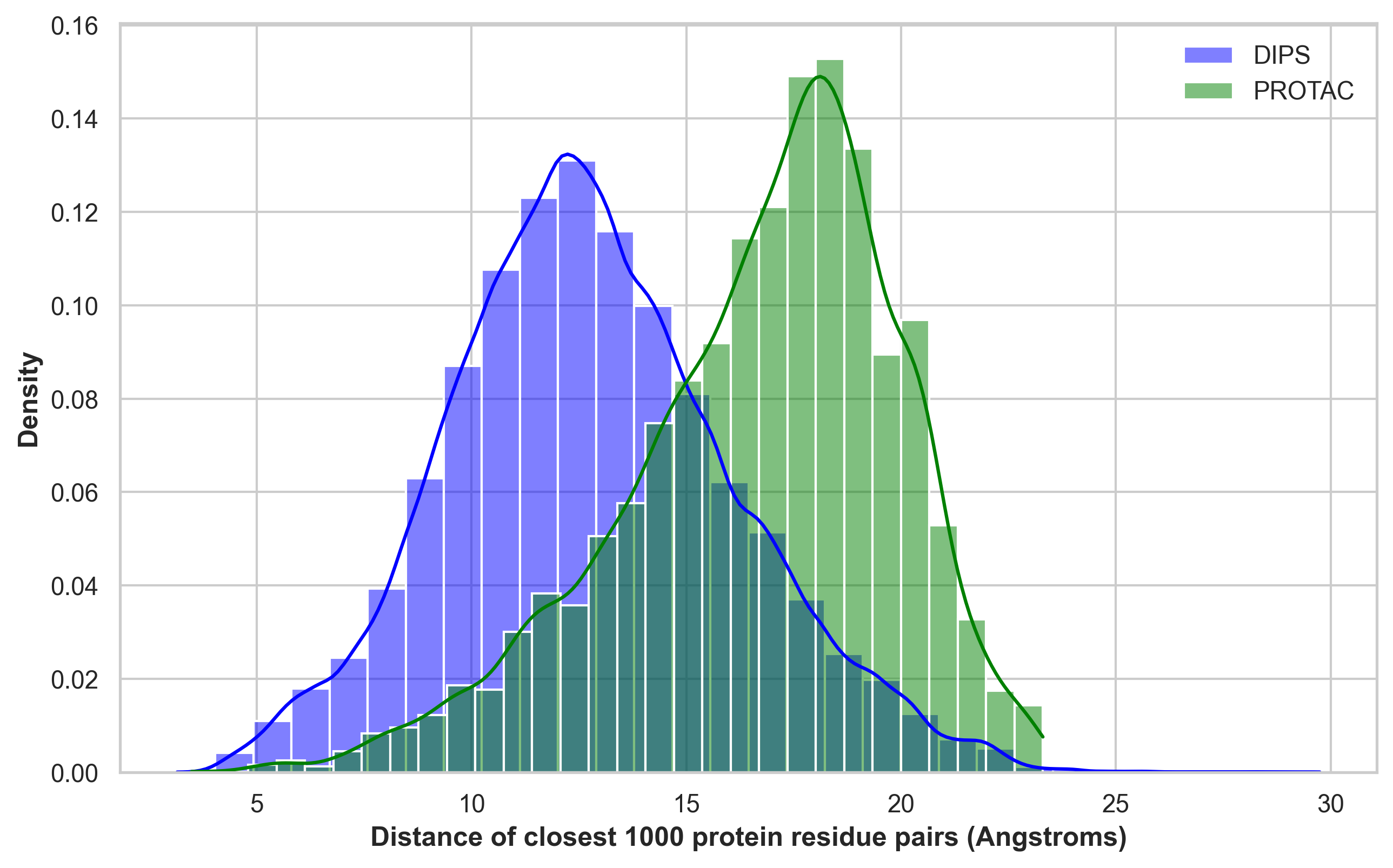}

\vskip -0.1in
\caption{Distribution of the residue distances for the 1,000 closest pairs, drawn from the protein-protein binary dataset (DIPS) and the PROTAC test set. 
}\label{Fig:ppi-distance-stats}
\vskip -0.2in
\end{figure}

\subsection{Warhead docking analysis}
\label{subsec:warhead-docking}

\begin{table*}[ht]
\setlength\tabcolsep{3 pt}
\centering
\caption{The results of RMSD and centroid distance in warhead-pocket docking experiments. We present the docking accuracy obtained from two datasets: the curated PROTAC-DB dataset and 13 PROTAC ternary structures sourced from PDB. The metrics are computed based on full warhead structures or anchor bonds.}\label{tab:dock_wh}
\begin{tabular}{cc|c|ccc}
\midrule
\multicolumn{1}{c|}{Dataset}                                                         & Test component        & Size of test set & Mean(\textup{\AA})~$\downarrow$  & Median(\textup{\AA})~$\downarrow$ & \textless{}2\textup{\AA}~$\uparrow$ \\ \midrule
\multicolumn{1}{c|}{PROTAC-DB}                                  & full warhead (RMSD)  & 2245             & 2.86  & 1.59   & 0.565         \\ \midrule
\multicolumn{1}{c|}{\multirow{3}{*}{ternary complexes}} & full warhead (RMSD)  & 26               & 2.51  & 0.962  & 0.714         \\
\multicolumn{1}{c|}{}                                           & anchor bond (RMSD)   & 26               & 1.94  & 0.774  & 0.750         \\
\multicolumn{1}{c|}{}                                           & anchor bond centroids & 26               & 0.769 & 0.27   & 0.893         \\ \midrule
\end{tabular}
\end{table*}\label{tab:dock_wh}
\begin{table*}[ht]
\centering
\setlength\tabcolsep{3 pt}
\caption{Experimental results for PROTAC ternary structure docking using ground truth structures of protein-warhead and Glide-docked structures of protein-warhead.}
\begin{tabular}{c|ccc|ccc|cc}
\midrule
                 & \multicolumn{3}{c|}{Complex RMSD(\textup{\AA})}                              & \multicolumn{3}{c|}{Interface RMSD(\textup{\AA})}   & \multicolumn{2}{c}{Fnat}                        \\
            Methods                & Mean~$\downarrow$           & Median~$\downarrow$           & \textless{}20~$\uparrow$  & Mean~$\downarrow$          & Median~$\downarrow$        & \textless{}10~$\uparrow$  &Mean~$\downarrow$  &  Median~$\downarrow$          \\ \midrule
Docked warheads       & 13.01 & 13.51 &  0.985          &    8.51      & 8.52         & 0.658      & 0.231 & 0.194               \\
True warhead-protein                 & 12.85          & 13.12                  & 0.975          & 8.20         & 8.16         & 0.740      &0.264 & 0.211             \\ \midrule
\end{tabular}
\end{table*}\label{tab:exp_dock}
\paragraph{Experimental setup} To demonstrate the capability of existing search-based software in accurately predicting the relative structures of protein-warhead for PROTAC ternary structure docking, we curated a dataset from PROTAC-DB~\citep{weng2021protac} and PDB for warhead docking experiments. Initially, we retrieved E3 ligase and POI structures from PDB using the UniProt ID of existing PROTACs from PROTAC-DB. Subsequently, we filtered out all monomer proteins lacking binding small-molecule ligands. Among the remaining small molecules binding to POI or E3-ligase, we retained data where the maximum common structure with PROTAC warheads accounted for more than 80\% of the ligands. Finally, we removed all solvent molecules from the structures, keeping only the chains binding with the ligands. 2245 pairs of proteins and warheads-like small molecules are collected after this procedure.

Because of this large amount of data retrieved from PDB, we could draw a conclusion that the pocket pockets on existing POIs and E3 ligases are fully explored. Even if certain targets are considered undruggable by traditional inhibitors, it is not inherently challenging to locate a binder within the PDB to establish a preliminary understanding of the binding pocket's approximate location. Therefore, we mainly focused on pocket-specific docking instead of blind docking. We employed the Schrodinger suite for evaluating docking on our curated dataset and 13 ternary PROTAC structures. We initially prepared each protein structure by preprocessing them with the Protein Preparation Wizard. This step involved refining the protein structure, adding hydrogen atoms, and assigning bond orders. Subsequently, the ligand structures were processed through LigPrep to generate appropriate ionization and tautomeric states. The receptor grid of size 20\textup{\AA}$\times$20\textup{\AA}$\times$20\textup{\AA} was generated around the active site using the Receptor Grid Generation tool. For docking simulations, we employed Glide, Schrödinger's molecular docking program, applying standard precision (SP) and extra precision (XP) protocols. The SP protocol was employed for initial virtual screening, followed by more accurate XP docking to refine the binding poses and rank the ligands.

\paragraph{Result and discussion} As presented in Table~\ref{tab:dock_wh}, the existing method demonstrates proficiency in predicting over half of the warhead-like structures within 2\textup{\AA}, with a median RMSD of 1.59\textup{\AA} after the molecular docking. This substantiates the accurate prediction of protein-warhead structures for a broad spectrum of candidate warhead structures suitable for PROTAC design. Subsequently, we evaluate the docking accuracy of the protein-warhead structures in our test set. The median RMSD for docking warheads is 0.962\textup{\AA}. Additionally, since \ours{} relies on the positions of anchor bonds to maintain linker structures, we assess the docking accuracy of anchor bonds. The median centroid distances between the predicted anchor bonds and ground truth anchor bonds are 0.27\textup{\AA}, indicating a precise prediction of anchor bond positions. Finally, we present the experimental results of \ours{} using docked warheads instead of ground truth in Table~\ref{tab:exp_dock}. Despite a slight decline in performance, the decrease in metrics is relatively small when compared to the variations observed when employing different methods.

\subsection{Ablation study}
\label{sec:ablation}
\subsubsection{Ablation on training dataset and sampling}
\begin{table*}[ht]
\centering
\setlength\tabcolsep{3 pt}
\caption{Ablation study results presenting the impact of removing part of training data and steric guidance in sampling in the implementation of \ours{}.}
\begin{tabular}{c|ccc|ccc|cc}
\midrule
                 & \multicolumn{3}{c|}{Complex RMSD(\textup{\AA})}                              & \multicolumn{3}{c|}{Interface RMSD(\textup{\AA})}   & \multicolumn{2}{c}{Fnat}                        \\
            Methods                & Mean~$\downarrow$           & Median~$\downarrow$           & \textless{}20~$\uparrow$  & Mean~$\downarrow$          & Median~$\downarrow$        & \textless{}10~$\uparrow$  &Mean~$\downarrow$  &  Median~$\downarrow$          \\ \midrule
DIPS only       & 13.10 & 13.24 &  0.981          &    7.90      & 7.82         & 0.690      & 0.301 & 0.262               \\
E3 ligase dataset only      & 13.12 & 13.18 &  0.992          &    8.48      & 8.60         & 0.750      & 0.231 & 0.194               \\
No steric guidance       & 14.33 & 14.67 &  0.911          &    8.81      & 9.12         & 0.688      & 0.241 & 0.189               \\
Full \ours{}                 & 12.85          & 13.12                  & 0.975          & 8.20         & 8.16         & 0.740      &0.264 & 0.211             \\ \midrule
\end{tabular}
\end{table*}\label{tab:abla_dataset_sample}
In this ablation study, we explored the impact of removing a portion of the training data and incorporating the steric guidance for sampling. Notably, when we only train \ours{} on DIPS or E3 ligase dataset, the model's performance in complex RMSD exhibited a reduction. Regarding the interface RMSD and Fnat, the model achieved better results by only using the DIPS dataset. Therefore, we can conclude that the general dataset, DIPS provides more information on protein-protein interaction on the surface level. In order to balance the performance variation on complex RMSD and interface RMSD, we choose to use both training data sets for our main experiment. Additionally, the incorporation of the steric guidance technique demonstrated effectiveness in boosting the performance from the full complex level. These findings underscore the importance of training data and sampling strategies.
\subsubsection{Baseline results with ranking and clustering}
In our main experimental results, we mainly focused on the docking accuracy of the sampled structures of \ours{} and baseline methods, so we excluded the postprocessing procedure of PROsettac. Here we downloaded the top-cluster structures and top-scored structures from the official PROsettac repository \footnote{https://github.com/LondonLab/PRosettaC} after the ranking and clustering. They experimented on 7 unique structures downloaded from the PDB. We provided the results in Table~\ref{table:prosettac-cluster}.
\begin{table*}[ht]
\label{table:prosettac-cluster}
\centering
\setlength\tabcolsep{3 pt}
\caption{Ablation study on comparing \ours{} with PRosettaC with clustering and ranking}\label{table:prosettac-cluster}
\begin{tabular}{c|ccc|ccc|c}
\midrule
                 & \multicolumn{3}{c|}{Complex RMSD(\textup{\AA})}                              & \multicolumn{3}{c|}{Interface RMSD(\textup{\AA})}   & \multicolumn{1}{c}{Fnat}                        \\
            Methods & Mean~$\downarrow$ &Med~$\uparrow$& \textless{}20~$\uparrow$  & Mean~$\downarrow$        &Med~$\uparrow$& \textless{}10~$\uparrow$  &Mean~$\downarrow$        \\ \midrule
PRosettaC(top1 cluster)       & 10.38 & 10.39 &  1.000          &    8.19      & 8.21         & 0.543      & 0.284 \\
PRosettaC(top20\% Rosetta score)                 & 10.23          & 10.30                  & 0.996          & 7.79         & 7.60         & 0.610      &0.283 \\ 
\ours{}(no ranking/clustering)       & 12.55 & 12.60 &  0.963          &    8.01      & 8.08         & 0.829      & 0.313      \\
\midrule
\end{tabular}
\end{table*}
\subsubsection{Stability of using different training objective}
\label{sec:stability}
\begin{figure*}[htbp]
	\centering
	\subfloat[The \textbf{Translation tangent loss} curve when training the model with loss objective defined on \textbf{SE(3) tangent space}]{\includegraphics[width=.31\linewidth]{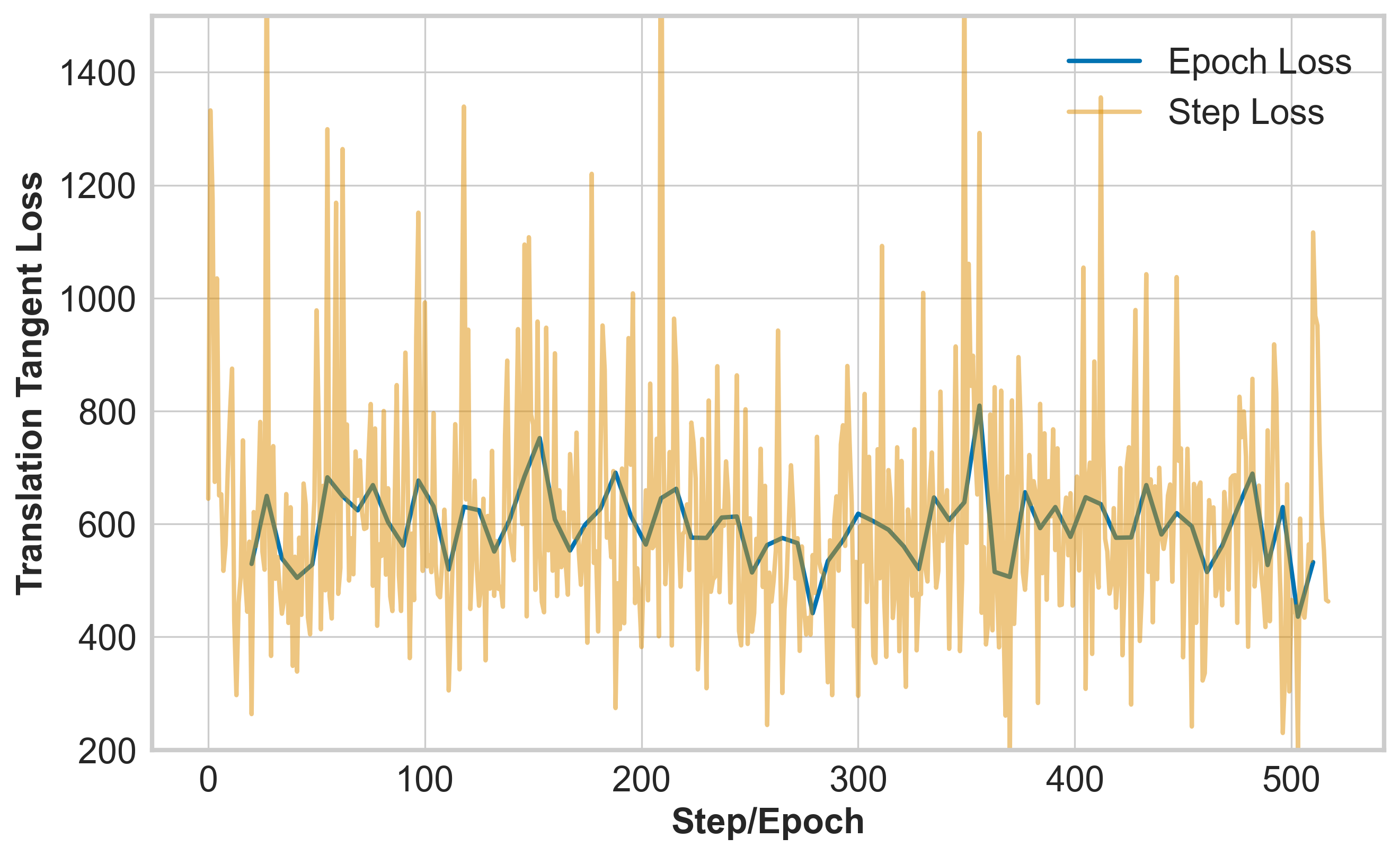}}\hspace{5pt}
        \subfloat[The \textbf{SO(3) tangent loss} curve when training the model with loss objective defined on \textbf{SE(3) tangent space}]{\includegraphics[width=.31\linewidth]{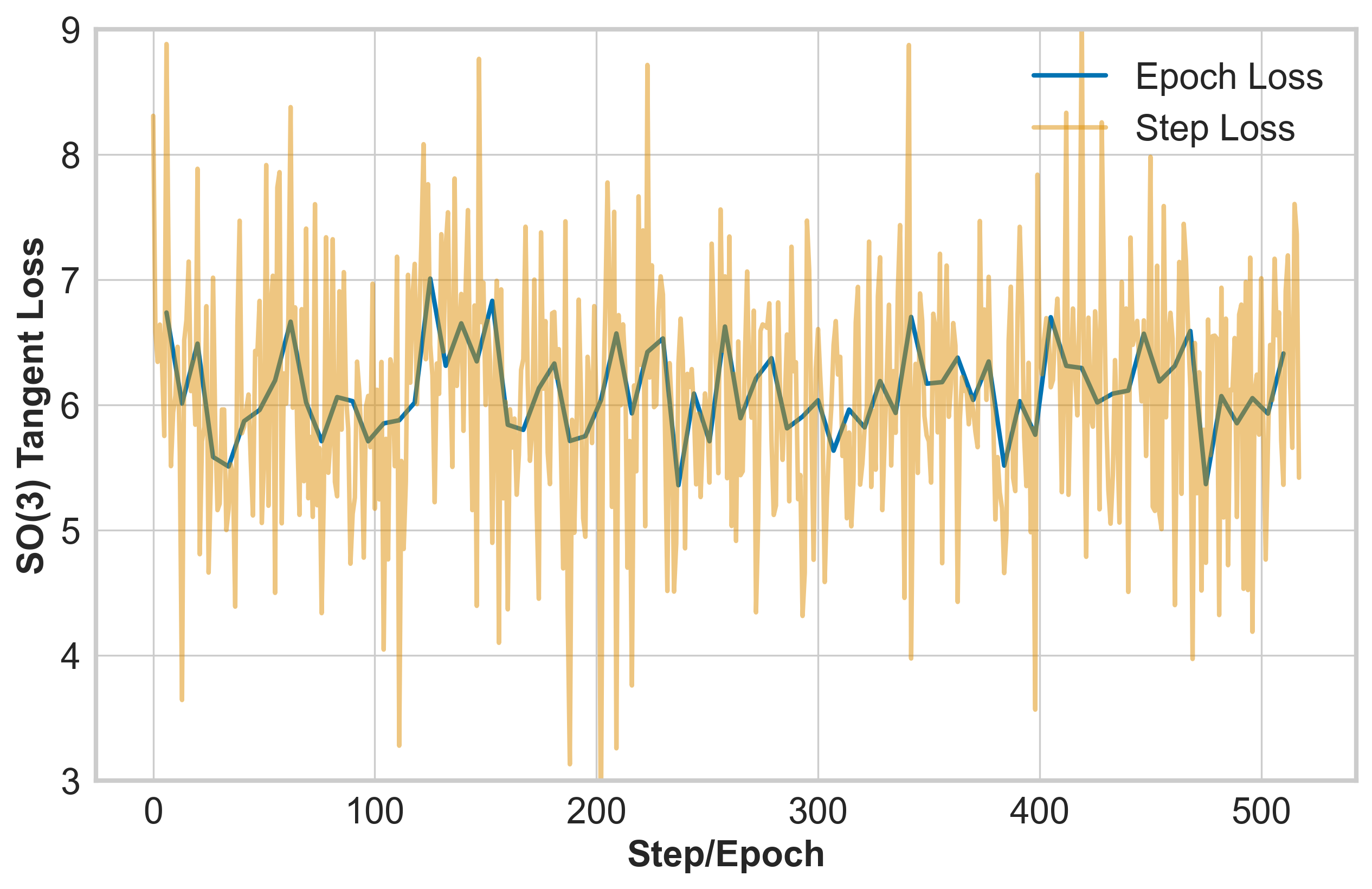}}\hspace{5pt}
	\subfloat[The \textbf{RMSD loss} curve when training the model with loss objective defined on \textbf{SE(3) tangent space}]{\includegraphics[width=.31\linewidth]{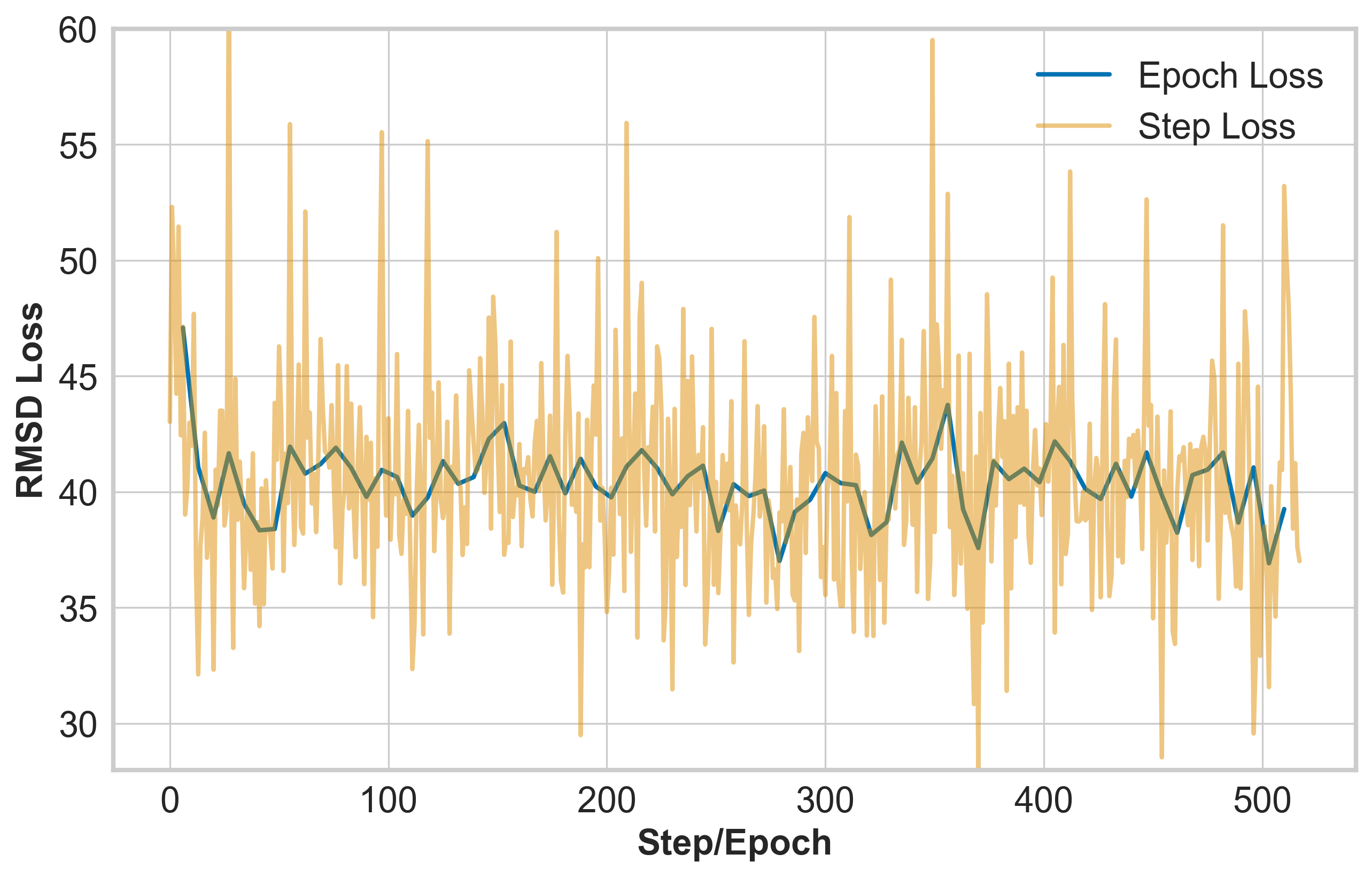}}\\
 \subfloat[The \textbf{Translation tangent loss} curve when training the model with \textbf{RMSD objective on the Euclidean space}]{\includegraphics[width=.31\linewidth]{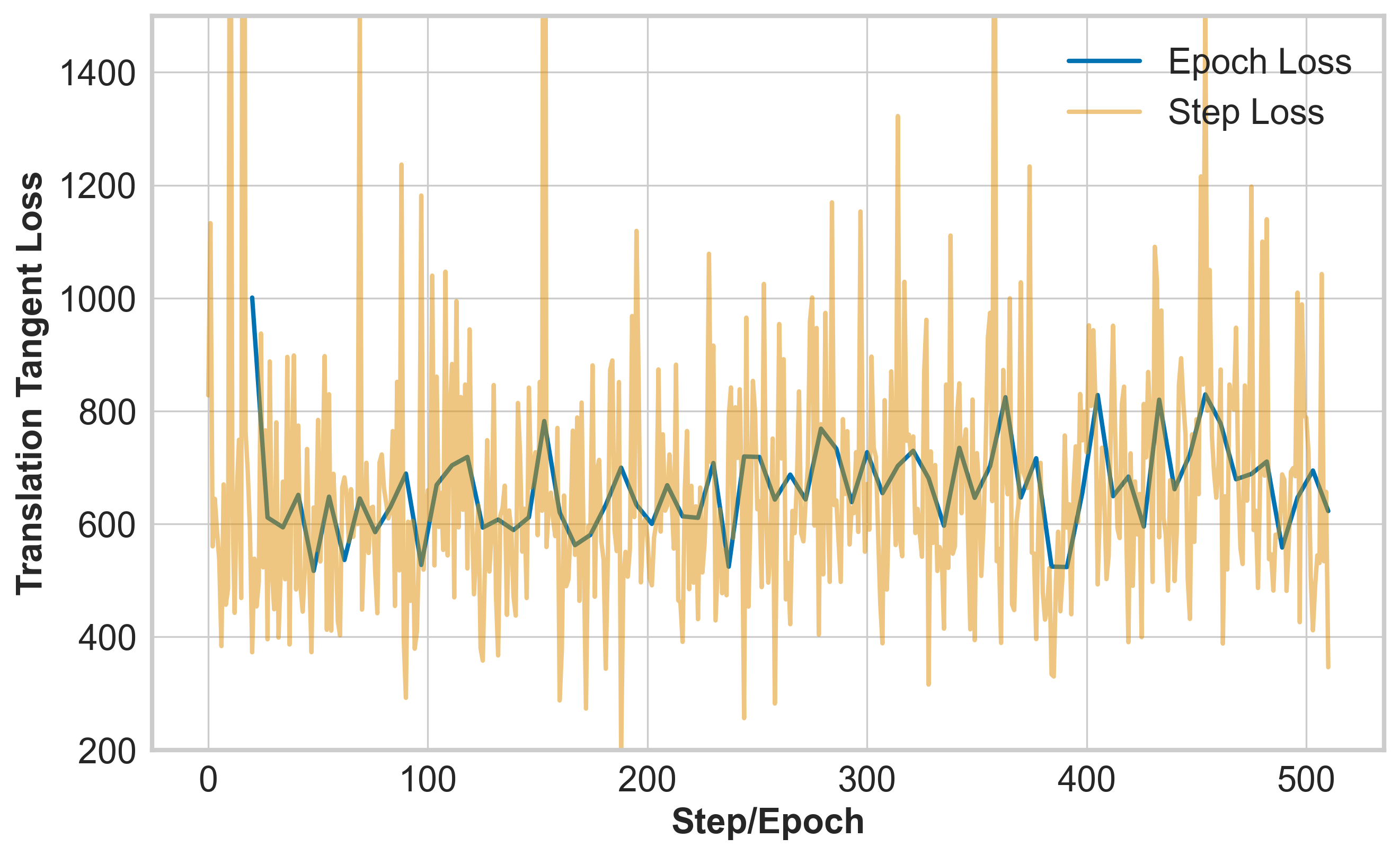}}\hspace{5pt}
        \subfloat[The \textbf{SO(3) tangent loss} curve when training the model with \textbf{RMSD objective on the Euclidean space}]{\includegraphics[width=.31\linewidth]{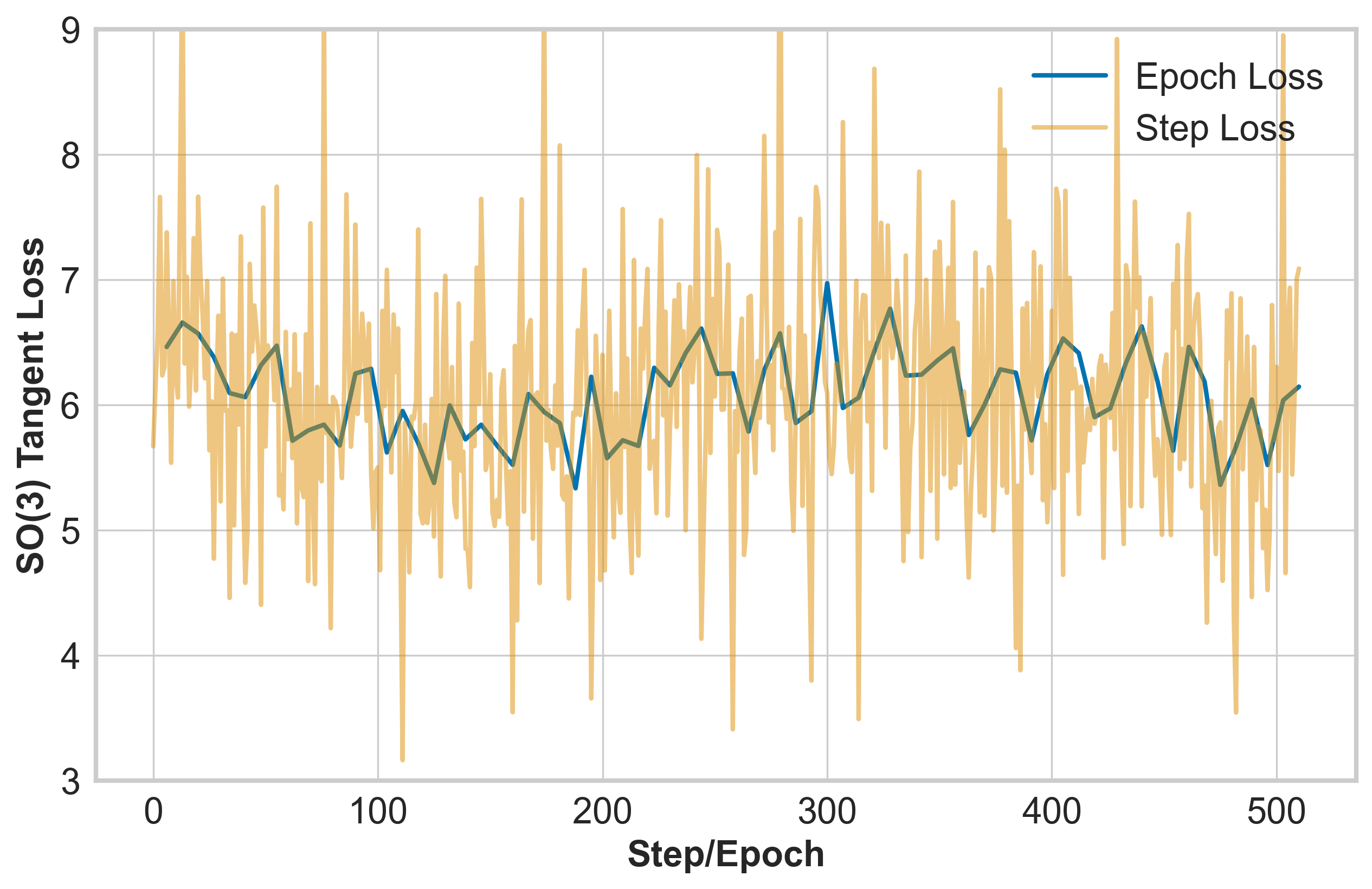}}\hspace{5pt}
	\subfloat[The \textbf{RMSD loss} curve when training the model with \textbf{RMSD objective on the Euclidean space}]{\includegraphics[width=.31\linewidth]{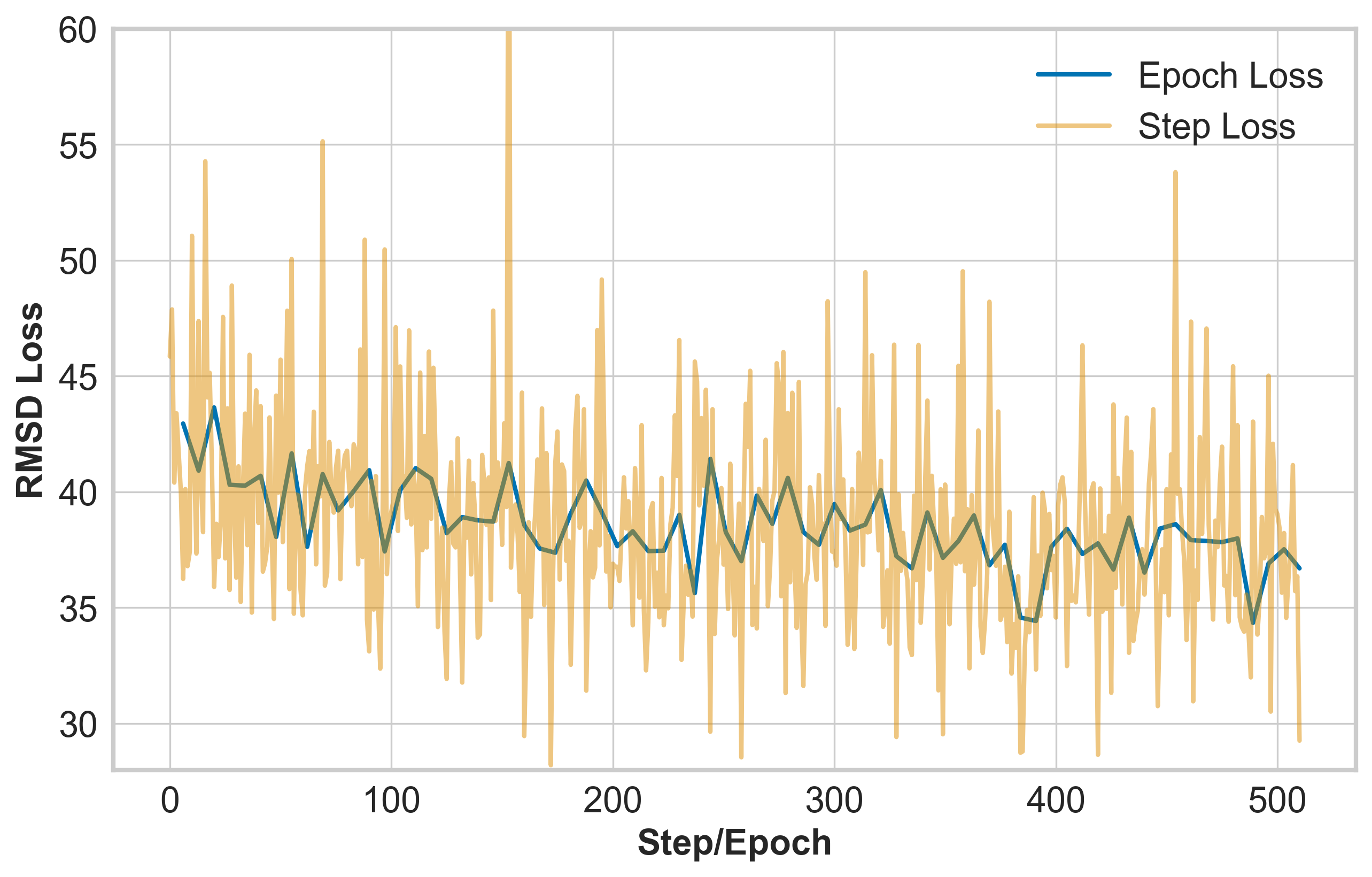}}
	\caption{The loss curve of \ours{} by using different training objective}\label{fig:train_curve}
\end{figure*}

In this section, we analyze the loss landscape (Figure~\ref{fig:train_curve}) corresponding to different training objectives. Figures~\ref{fig:train_curve}(a), (b), and (c) depict the training loss curves for employing the training objective within the tangent space of $\text{SO(3)} \times \mathbb{R}^3$, as previous work by \citet{yim2023fast}. The translation loss in $\mathbb{R}^3$ and rotation loss in SO(3) exhibit noise, making it challenging to monitor their convergence. Figures~\ref{fig:train_curve}(d), (e), and (f) depict the training loss curves for employing our RMSD loss. Conversely, the loss curve in Figure~\ref{fig:train_curve}(f) demonstrates more stable convergence in comparison to the conventional equivariant flow matching loss for rotation and translation. Simultaneously, when utilizing the RMSD loss, the translation and rotation losses continue to converge within the noisy landscape.

From a theoretical point of view, the training objective could be defined as the distance measure in three different spaces including Euclidean space $\mathbb{R}^{3\times N}$, the SE(3) manifold, and linker constraint space $\mathcal{M}_\ell$. They are hierarchically related as subspaces, which means that $ \mathcal{M}_\ell \subset \text{SE(3)} \subset \mathbb{R}^{3\times N}$. The RMSD loss we use corresponds to the distance in Euclidean space, where optimizing it is equivalent to optimizing a weighted sum of the distances in $\mathbb{R}^{3}$ and SO(3). The weight is automatically adjusted by the inner distances of E3 ligase coordinates.

Traditional CFM loss treats SE(3) as the product space of $\mathbb{R}^{3}$ and SO(3) where the training objective is separated. From the above loss analysis, dividing SE(3) into translation and rotation is hard for the optimizer. 
Our training objective draws inspiration from prior work~\citep{jing2023alphafold}, which transitions the flow matching loss from the space \(\mathbb{R}^{3\times N}\) to the quotient space \(\mathbb{R}^{3\times N} / SE(3)\). 
In \ours{}, the loss is lifted into $\mathbb{R}^{3n}$ to jointly optimize the loss objective on translation space and rotation space that demonstrates practical efficacy.

\end{document}